\documentclass{ametsocV6.1}

\let\internallinenumbers\relax

\usepackage{multirow}

\title{Dependence of the precipitation intensity distribution on spatial and temporal resolution in observations}

\authors{Akshay Rajeev,\aff{a}\correspondingauthor{Akshay Rajeev, ar942@cornell.edu} Maxim A. Masleyev,\aff{a} and Angeline G. Pendergrass\aff{a}}

\affiliation{\aff{a}{Department of Earth and Atmospheric Sciences, Cornell University, Ithaca, NY, USA}}

\begin{document}

\abstract{Precipitation intensity varies with both spatial and temporal resolution, with broad implications across climate science, yet the nature of this dependence is poorly characterized. We investigate how precipitation intensity distributions change with spatial and temporal resolutions and introduce a framework that approximates these changes between resolutions. Using merged satellite-gauge data regridded to various grid sizes, we construct precipitation amount distributions for spatial resolutions ranging from a base resolution of 0.1º for IMERG and 1º for the FROGS database to a coarse 10º. Similarly, we use gauge observations coarsened to temporal resolutions ranging from 1-minute to 24-hours to construct precipitation amount distributions. This approach allows us to remove physical differences in order to isolate the effect of resolution on precipitation. Results show that the precipitation intensity distribution varies across spatial and temporal resolutions even without changes in physical processes. We demonstrate that these resolution-dependent changes in the precipitation amount distribution can be effectively characterized by shifting the distribution in logarithmic space without changing its shape, called the shift mode. Our findings emphasize the need to account for the effects of scale on precipitation when comparing and interpreting differences in gridded precipitation from models or observations.}

\maketitle

%
%
%
\statement
The amount of rainfall measured during a storm often depends on the level of detail provided by the observational dataset or model used. As a result, rainfall intensity estimates can differ substantially between different datasets, making it hard to know which data to trust. It has been unclear whether these differences reflect real physical processes being captured at finer detail, or simply an inherent effect of resolution itself. Using satellite and gauge precipitation datasets, we examined how rainfall varies across spatial and temporal resolutions using the same observation and without altering any physical processes. We find that the resolution dependence of rainfall is predictable, as it occurs even when the physical processes are unchanged. 
%
%
%

%
\section{Introduction}
\label{sec1}
Observations and climate models operate across a range of scales and it has been found that precipitation properties are sensitive to the spatial and temporal resolution at which they are examined \citep{eggert_temporal_2015, chen_precipitation_2019,martinez-villalobos_why_2019}. This sensitivity has critical consequences for disaster planning, water management, agriculture, and infrastructure development, which require precipitation information at various spatial and temporal resolutions. Understanding how precipitation intensity varies with resolution is fundamental to improving our interpretation of atmospheric processes and its application. 

Precipitation, like various other atmospheric variables, has been viewed using the scale-bound framework and the scale-invariant framework. The scale-bound framework is the idea that atmospheric phenomena at different scales (synoptic, mesoscale and convective) are governed by different independent physical processes, arguing that there is no single scaling relationship connecting different scales \citep{mitchell_overview_1976}. When applied to precipitation, this framework suggests that precipitation from large stratiform structures and localized convective structures are governed by different processes. Despite a lack of empirical evidence, aspects of atmospheric and climate science still operate within this scale-bound framework \citep{lovejoy_review_2023}, which is evident in much of the literature using weather and climate models to examine atmospheric processes. For example, it has been asserted that the increased intensity of simulated precipitation at higher spatial resolution is due to improvements in the representation of physical processes \citep{kopparla_improved_2013,slingo_ambitious_2022, jong_increases_2023, lin_resolution_2023}, despite that model evaluation studies do not uniformly find this \citep{bador_impact_2020,nishant_impact_2022}. In the former case, the approach to understand how spatial resolution affects precipitation intensity adopts the scale-bound framework, in which different resolutions are assumed to be governed by distinct physical processes, and changes in intensity are attributed to improvements in the representation of specific processes at a given resolution. However, the underlying equations that govern these models scale in space and time follow the scale-invariant framework \citep{lovejoy_review_2023}.

The scale-invariant framework suggests that different structures can emerge from a single scale-invariant relation \citep{schertzer_generalised_1985}, and was investigated primarily using the random cascades theory \citep{mandelbrot_intermittent_1974}. Early work following this scale-invariant framework used multifractal relations, establishing that precipitation has anisotropic scaling structures and describe how precipitation is redistributed between different scales \citep{lovejoy_generalized_1985}. In the case of spatial resolution, this redistribution is uneven and intermittent, with some grid cells getting intense precipitation while most receive almost none. Subsequent studies further demonstrated that precipitation cascades across spatial and temporal scales using various observational datasets - precipitation gauge networks \citep{fraedrich_scaling_1993, olsson_effect_1999,marani_nonpowerlawscale_2005}, radar datasets \citep{over_statistical_1994,over_spacetime_1996,venugopal_evidence_1999,deidda_rainfall_2000,veneziano_imperfect_2006}, and satellite products \citep{lovejoy_remarkable_2008, gebremichael_scaling_2008}.

Evidence supporting the scale-invariant framework has also been documented for clouds, where cloud size distributions follow a power-law relationship \citep{wood_distribution_2011,lovejoy_review_2023,dewitt_climatologically_2024}. Precipitation, like clouds, exhibits scale-invariant structural properties; however, unlike clouds, scale-invariance also governs precipitation intensity, such that spatial and temporal resolution can systematically influence intensity \citep{fraedrich_scaling_1993,olsson_effect_1999,deidda_rainfall_2000}. While these studies characterize how precipitation scaling exponents and cascade parameters vary across scales, they do not directly quantify the difference in precipitation intensity between resolutions. Separately, studies not employing this scaling framework also provide evidence for such resolution dependence in precipitation. These studies found precipitation frequency and intensity to be sensitive to both the spatial and temporal resolutions, with the intensity decreasing at coarser resolutions \citep{tan_performance_2017,chen_dependence_2018,chen_precipitation_2019}. Further evidence has arisen from studies that argue that hourly precipitation exhibits fundamentally different characteristics than daily averages \citep{trenberth_intermittency_2017, trenberth_nearglobal_2018}. Subsequently, \citet{martinez-villalobos_why_2019} characterized the differences in precipitation intensity distributions across different aggregation periods while connecting them to underlying physical processes, though they address only the temporal aspect. No existing framework provides a unified approach to quantifying differences in precipitation intensity between resolutions while retaining information about the scale of the precipitating system which can be applied to both spatial and temporal resolutions. 

Building on this understanding of the scale invariance of precipitation, we develop a framework to characterize differences in precipitation intensity between resolutions based on multiple observational datasets. In addition to quantifying resolution dependent variations in precipitation intensity, the framework also provides information about spatial and temporal scale of the precipitation. First, we examine the redistribution of precipitation, in terms of its intensity, between different spatial resolutions using observational precipitation datasets regridded to multiple grid sizes. This approach holds the underlying physics constant, isolating the effect of resolution. Subsequently, we characterize the effect of spatial and temporal resolution on precipitation amount distribution by shifting the distribution while holding its shape fixed, called the shift mode \citep[adapted from][]{pendergrass_two_2014}. We apply this framework to multiple observational datasets to understand the influence of spatial resolution on precipitation intensity. Finally, we also use the framework to quantify changes in the intensity distribution at different timescales in an observational dataset with very high (1-minute) native temporal resolution.

\section{Data and methods}
\label{sec2}
\subsection{Spatial resolution dependence: Merged satellite-gauge observations}\label{sec2a}
This study primarily uses precipitation estimates from the operational quasi-global Integrated Multi-satellitE Retrievals for GPM (IMERG v07B) to understand the dependence of precipitation on spatial resolution \citep{g.j.huffman_gpm_2023}. The IMERG algorithm merges multiple satellite microwave precipitation estimates with microwave-calibrated infrared estimates, gauge analyses and other precipitation estimators at fine spatio-temporal scales for the entire globe. We use the final run of the satellite product, ranging from 1998 to 2024, at daily time scales with a native spatial resolution of 0.1° \citep{g.j.huffman_gpm_2023}. The IMERG Final run incorporates monthly gauge analysis to remove biases. For merged products like IMERG, the effective resolution determines the highest spatio-temporal scale to which a gridded product can resolve precipitation patterns. The effective resolution depends on the instrument spatial resolution, merging algorithm, temporal resolution of the product, location, precipitation type and season. Over CONUS, the effective resolution of IMERG at daily timescales ranges between 20-80 km \citep{guilloteau_multiscale_2020}.

We also use observations from the Frequent Rainfall Observations on GridS (FROGS) database to test the applicability of the method to other datasets \citep{roca_frogs_2019}. The FROGS database consists of gridded daily precipitation products regridded to a common 1º X 1º grid. We focus on a few merged satellite datasets which we select by prioritizing overlap of temporal range, 2002-2020, and spatial domain of IMERG, 60ºS-60ºN. The Global Precipitation Climatology Project (GPCP) version 3.2 provides global estimates of surface precipitation by merging low earth orbit passive microwave (PMW) estimates, infrared data from geosynchronous satellites (GEO-IR), sounders and surface precipitation gauges \citep{huffman_new_2023}. While the GPCP v3.2 daily product incorporates half-hourly estimates from IMERG version 6 Final product, the former focuses on long-term homogeneity compared to grid-scale accuracy in IMERG \citep{huffman_integrated_2020}. The Climate Prediction Center Morphing Technique with corrected bias (CMORPH-CRT, version 1) provides satellite precipitation estimates by merging level-2 PMW retrievals of precipitation rates, GEO-IR estimates, and satellite-based snow/ice maps \citep{xie_reprocessed_2017}. CMORPH-CRT is calibrated over land using collocated CPC gauge analysis and over the ocean using GPCP pentad merged analysis. Similarly, another merged satellite product from FROGS that is used here is the Global Interpolated RAinFall Estimation (GIRAFE), which takes precipitation retrievals from PMW and GEO-IR satellite data as the input \citep{niedorf_girafe_2024}. Finally, the Precipitation Estimation from Remotely Sensed Information using Artificial Neural Networks–Climate Data Record (PERSIANN-CDR v1 r1) uses GridSat-B1 infrared data and the PERSIANN algorithm to generate a consistent, long-term and global precipitation dataset for studying trends in daily precipitation \citep{ashouri_persianncdr_2015}. It also utilizes the GPCP monthly product to maintain climatological consistency. In all of the merged products listed above, there is considerable overlap in the satellite constellation used, and most of the differences probably arise from their algorithmic design.

\subsection{Temporal resolution dependence: Very high temporal resolution gauge observations}
To assess the dependence of precipitation on temporal resolution, we use in-situ observations from the US Department of Energy’s Atmospheric Radiation Measurement (ARM) program (Stokes and Schwartz, 1994). We chose measurements from ARM long-term observational sites: one mid-latitude site (36.61ºN), the Southern Great Plains (SGP) at Lamont, Oklahoma \citep{zhu_weighing_2026} and one tropical site (2.06ºS), the Tropical West Pacific C1 facility (TWP) at Manus Island \citep{kyrouac_surface_2026}. The difference in the location of the sites at different latitudes enables us to examine the consistency of the our framework across precipitation regimes. These observations are constrained by logistical and environmental factors, such as sensor maintenance and coverage gaps due to which the temporal range of the two observations differ. For the SGP site, we use precipitation rates (mm/hr) from November 2016 to March 2024 while, data from the TWP site are from November 2004 to July 2014.

\subsection{Low-resolution generation via coarsening and construction of precipitation amount distributions}
\begin{figure*}
 \centerline{\includegraphics[width=6.5in,angle=0]{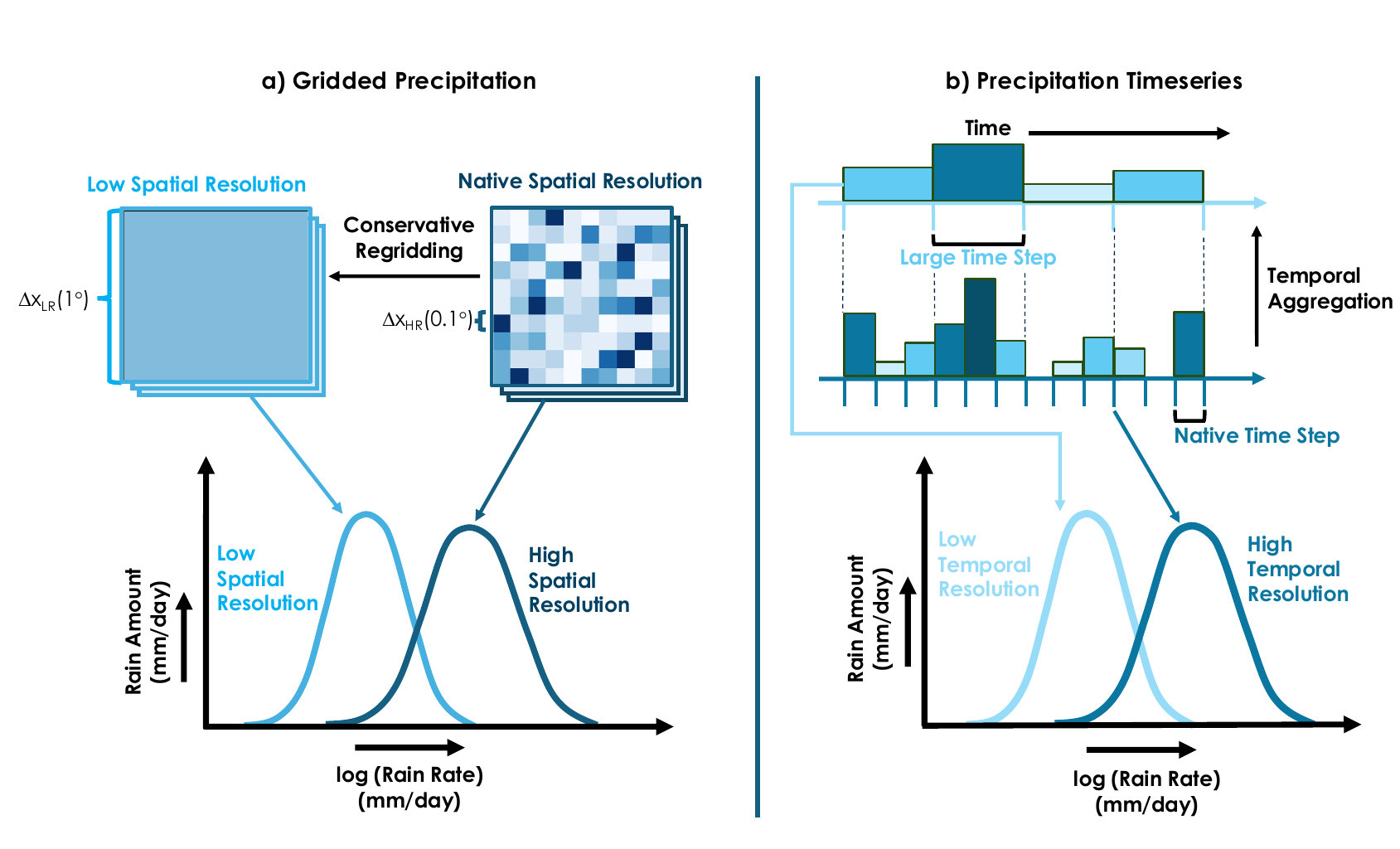}}
\caption{Schematic explaining the data preprocessing and construction of precipitation amount distributions. a) Data preprocessing for spatial scaling, where the native grid is coarsened to lower spatial resolution. For gridded precipitation each spatial resolution, the precipitation amount distribution is constructed at each grid point and averaged over the entire domain. Here, the native spatial resolution data is 0.1º and the low spatial resolution is 1º. b) Data preprocessing for temporal scaling, where the native time step is aggregated to lower temporal resolutions. Precipitation amount distribution is constructed for precipitation timeseries at each temporal resolution.} \label{fig1}
\end{figure*}

To understand the dependence of precipitation on spatial resolution, we regrid the data to coarser spatial resolution and construct precipitation amount distribution for each grid sizes (Fig.~\ref{fig1}a). We use conservative regridding to coarsen the native 0.1º IMERG product to 0.25º, 1º, 2º, 5º and 10º. Conservative regridding preserves the integral of precipitation amount across grids of different resolutions \citep{jones_first_1999}, ensuring that the underlying physics remains constant across different spatial resolutions. Similarly, we coarsen each dataset in the FROGS database from its standard 1º horizontal resolution to the 2º, 5º and 10º.

\begin{figure*}
\noindent\includegraphics[width=5.5in,angle=0]{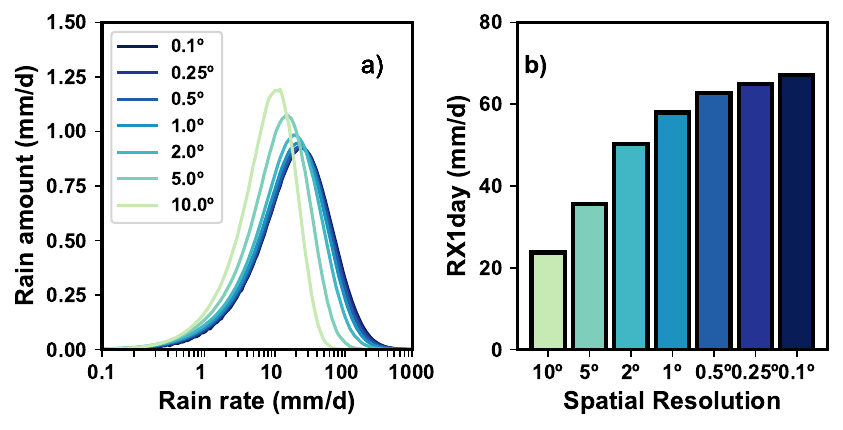}\\
\caption{a) Precipitation amount distributions at different spatial resolutions, b) Globally averaged RX1Day values calculated for different spatial resolutions. Each distribution was created from the native (0.1º) and regridded (0.25º,0.5º,1º,2º,5º and 10º) spatial resolutions from IMERG. The different colors correspond to different spatial resolutions.}\label{fig2}
\end{figure*}

For each spatial resolution and dataset, we then calculate the precipitation amount distribution as shown in Fig.~\ref{fig2}a and Fig.S1, which enables cross-resolution comparison of precipitation by collapsing spatial information while preserving precipitation properties. The precipitation amount distribution is the total accumulated precipitation at each rain rate bin normalized by the total number of days, i.e. it describes how much precipitation falls at different precipitation rates and is constructed with logarithmically-spaced bins \citep{pendergrass_two_2014}. Because of this logarithmic spacing, movement along this axis signifies a percentage change in precipitation rate. We construct the precipitation amount distribution at each grid point by estimating the total amount of precipitation that falls in each precipitation rate bin. We then calculate the area-weighted average over all grid points to generate the globally aggregated distribution. A mathematical description of the calculation of the precipitation amount distribution is provided in \citet{pendergrass_two_2014}. It should be noted that in this study, we use precipitation rate and intensity interchangeably to refer to precipitation accumulation per unit time (mm/day and mm/min) at the specified spatial and temporal resolution.

\begin{figure*}
 \centerline{\includegraphics[width=6.5in,angle=0]{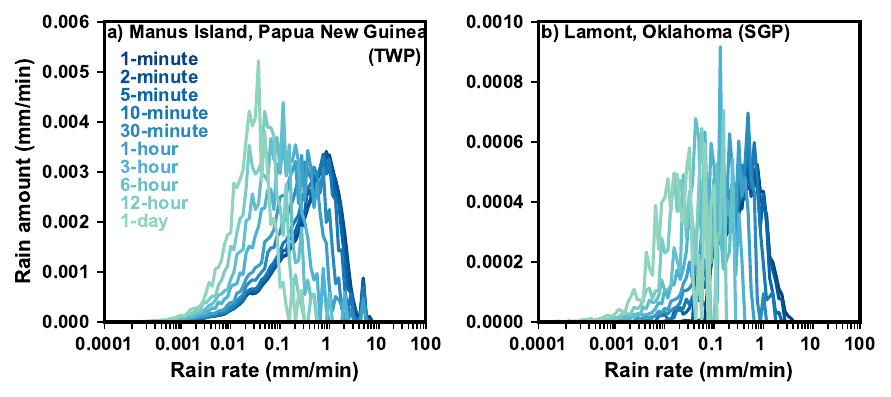}}
\caption{ Precipitation amount distributions at different temporal resolutions from DOE ARM gauge data at a) Manus Island (TWP) and b) Lamont, Oklahoma (SGP). Each distribution is created from the native (1-minute) and aggregated (2-minute, 5-minute, 10-minute, 30-minute, 1-hour, 3-hour, 6-hour, 12-hour and 1-day) gauge data.}\label{fig3}
\end{figure*}

For the application to temporal resolution with ARM gauge observations at 1-minute temporal resolution, we first remove days with missing data as it may propagate errors during resampling. The lengths of the resulting timeseries are 3515 days at TWP and 1768 days at SGP. Analogously to spatial coarsening, we coarsen the data in time (Fig.~\ref{fig1}b). We then calculate precipitation amount distributions for 1-minute, 2-minute, 5-minute, 10-minute, 30-minute, 1-hour, 3-hour, 6-hour, 12-hour and daily resolutions (Fig.~\ref{fig3}). The construction of these distribution for each temporal resolution enables us to compare changes in precipitation due to the length of accumulation.

\subsection{Joint distribution of precipitation amount between resolution pairs}
To develop a thorough understanding of the dependence of precipitation on resolution, we analyze how precipitation behaves on the joint space of high- and low-resolution intensities. To accomplish this, we introduce the joint precipitation amount distribution between high and low resolutions. This distribution is analogous to a joint histogram, but instead of precipitation frequency, it is based on precipitation amount. The joint precipitation amount distribution describes the contribution of each pair of coincident high- and low-spatial resolution rain rates to the total precipitation amount.

The calculation of the joint amount distribution requires matching, overlapping grids for each high- and low-spatial resolution pair. Having constructed the coarse resolution grids so that no high-resolution grids straddle multiple low-resolution grid cells, we match each high-resolution grid cell with the low-resolution grid cell in which it is contained. For each overlapping grid-cell pair, we calculate the two-dimensional histogram of co-occurring precipitation rates from the high- and low-resolution datasets, where \(R_H^c\) and \(R_L^c\) denote the high- (\(H\)) and low- (\(L\)) resolution precipitation rate axis. Similarly, to calculate the joint amount distributions between temporal resolution pairs, we calculate the two-dimensional histogram of co-occurring precipitation rates from the high- and low-resolution timeseries. 

Unlike a joint histogram, where the frequency fully describes the space, the joint precipitation amount distribution can be summed for both the high- and low-resolution data with marginal totals that can differ. For each bin, summing the total precipitation from the high-resolution data yields the high-resolution joint distribution, which is then normalized by the total number of days. If we instead sum the low-resolution precipitation, this forms the low-resolution joint amount distribution, which is also normalized by the total number of days. In the case of spatial resolution, we generate the globally-aggregated joint distributions by calculating the area-weighted average over all grid points. We indicate the high-resolution joint precipitation amount distribution with \(Z_H(R_H^c,R_L^c)\), and the low-resolution joint distribution with \(Z_L(R_H^c,R_L^c)\). A mathematical description of the calculation of the precipitation amount distribution is provided in the Appendix.

The joint distributions \(Z_H(R_H^c,R_L^c)\) and \(Z_L(R_H^c,R_L^c)\) are each related to the total high- and low-resolution precipitation amount distributions (\(P_H\) and \(P_L\)). Summing across the high-resolution rain rate axis (\(R_H^c\)) in \(Z_H(R_H^c,R_L^c)\) recovers the low-resolution total precipitation amount distribution, \(\sum_{R_H^c}Z_H(R_H^c,R_L^c) = P_L\). Similarly, summing across the low-resolution precipitation rate axis (\(R_L^c\)) in \(Z_H\) recovers the high-resolution total precipitation amount distribution, \(\sum_{R_L^c}Z_H(R_H^c,R_L^c) = P_H \). Additionally, summing across \(R_H^c\) over \(Z_L\) recovers \(P_L\), \(\sum_{H}Z_L(R_H^c,R_L^c) = P_L \). However, summing across \(R_L^c\) does not result in \(P_H\) except in the special case that \(Z_L(R_H^c,R_L^c)= Z_H(R_H^c,R_L^c)\). 

\subsection{Directly-calculated shift mode}
Using the precipitation amount distributions at different spatial and temporal resolutions described in Section~\ref{sec2}a-c, we can quantify the effect of changing resolution on the precipitation amount distribution. To achieve this, we build on the framework of shift and increase modes of change described in \citet{pendergrass_two_2014}. Because the total area-averaged precipitation does not change with resolution, due to our choice of construction (see Section ~\ref{sec2}c), the increase mode is unnecessary here. While \citet{pendergrass_two_2014} calculates the magnitude of the modes using the closest possible fit to the change between distributions, this relies on the changes being small. Because this criteria is not always met for the resolution-pairs we consider, we instead use a method of direct calculation to quantify the shift mode. Using the fractional change in the rate above (or below) which half of precipitation falls as a measure of the precipitation amount peak of the amount distribution, we define the directly-calculated shift mode as the change in the location of the precipitation amount peak,
\begin{equation} \label{eq:1}
b_o=\frac{r_{1/2}^{H} - r_{1/2}^{L}}{r_{1/2}^{L}},
\end{equation}
\noindent where \(r_{1/2}^{H}\) and \(r_{1/2}^{L}\) are the precipitation rates corresponding to the \(50^{th}\) percentile precipitation for the high- and low-resolution precipitation amount distributions respectively. To avoid confusion, we refer to the shift mode based on the best fit as the fitted shift mode (\(b\)) and that from Equation~\ref{eq:1} as the directly-calculated shift mode (\(b_o\)). Like the fitted shift mode, the directly-calculated shift mode is a unitless fraction, or can be expressed in percentage. We explore the relationship between the fitted and directly-calculated shift modes when applied to the quantification of differences in precipitation intensity between spatial resolution in Section~\ref{sec3}e.

After calculating the shift mode \(b_o\), we shift the low-resolution distribution along the log-rate axis using \(b_o\) as the offset. Since this offset simply shifts each point on the distribution by the same magnitude, the transformation does not change the shape of the low-resolution distribution. The shifted low-resolution distribution helps us quantify how well \(b_o\) captures the difference between two resolutions and is discussed in Section~\ref{sec3}d. 

\section{Dependence of precipitation intensity on spatial resolution}
\label{sec3}

We first focus on characterizing how precipitation intensity changes between spatial resolutions. For this we use the globally-aggregated precipitation amount distributions for IMERG at different spatial resolutions (Fig.~\ref{fig2}a). As a proxy for the centroid of the amount distribution, we use the precipitation amount peak \citep{pendergrass_climatological_2017,kooperman_robust_2016}, which we approximate as the rain rate above which the half of precipitation falls, \(r_{1/2}\), is 19 mm/day for the 0.1º spatial resolution. The amount peak moves to 18.37 mm/day at 0.25º, 17.70 mm/day at 0.5º, 16.38 mm/day at 1º and 10.85 mm/day at 5º spatial resolution. The movement to lower rain rates with decreasing spatial resolution is not limited to the amount peak, but extends to the entire distribution. Furthermore, the shape of the amount distribution is nearly fixed for small resolution differences, particularly for changes in grid sizes less than 1º. The movement of the precipitation amount distribution to higher intensities indicates that there is more precipitation falling at heavier rates, whereas the amount of precipitation falling at lighter precipitation rates decreases with increasing spatial resolution. In addition to moving to higher intensities, the precipitation amount distribution also widens with increasing resolution, though this is only substantial for large differences in spatial resolution. The spatial-resolution-dependent changes in the precipitation amount distribution found here do not reflect differences in underlying physics, which is fixed by our choice of construction (see Section ~\ref{sec2}c). While the precipitation frequency distribution (Fig.S2) also contains such resolution dependent differences, they do not exhibit as a shift along the precipitation rate axis. Instead, precipitation frequency increases with spatial resolution at higher precipitation rates, while it decreases at lower intensities. We next investigate how these resolution-dependent changes manifest in the upper tail of the precipitation distribution.

\subsection{Dependence of extreme precipitation on spatial resolution}
Next, we examine the effect of the change in spatial resolution on extreme precipitation using RX1day, which is the maximum 1-day precipitation amount each year \citep{alexander_global_2006,alexander_use_2019}. We calculate RX1day at each grid point and then take the global average, for each spatial resolution (Fig.~\ref{fig2}b). Rx1day globally averaged at the native 0.1º resolution of IMERG is 67 mm/day, which reduces to 64 mm/day at 0.25º, 57 mm/day at 1º and 35.7 mm/day at 5º. This decrease in Rx1day with decreasing resolution indicates that the intensity of extreme precipitation increases with spatial resolution. Further discussion focusing on the resolution dependence of extreme precipitation in the context of broader literature is provided in Section~\ref{sec5}.

\subsection{Application of joint precipitation amount distributions to changing spatial resolution}

\begin{figure*}
 \centerline{\includegraphics[width=6.5in,angle=0]{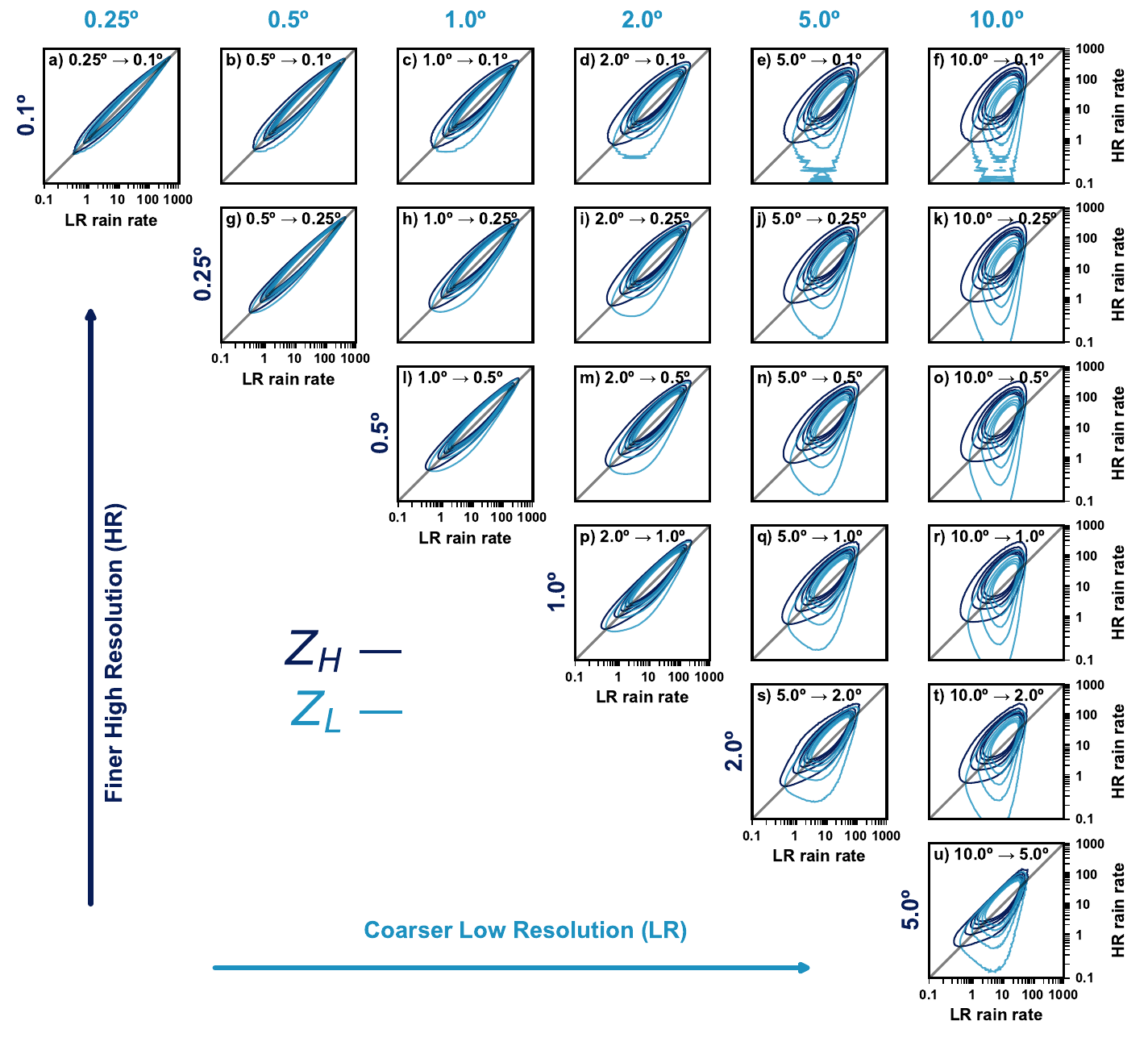}}
\caption{Joint precipitation amount distributions between high and low spatial resolutions for IMERG coarsened to different resolutions. In each panel, the joint precipitation amount for high-resolution is depicted in dark blue and the joint precipitation amount for low-resolution distributions are depicted in light blue. The high-resolution datasets used are 0.1º, 0.25º,0.5º, 1º, 2º and 5º and the low-resolution datasets are 0.25º,0.5º, 1º, 2º, 5º and 10º. Each row corresponds to a fixed high-resolution, and each column corresponds to a fixed low-resolution. Precipitation rate bins from the low-resolution data are plotted along the $x$-axis, and those from the high-resolution data are along the $y$-axis.}\label{fig4}
\end{figure*}

Next, we examine the joint precipitation amount distribution to investigate how the precipitation distribution varies between spatial resolution pairs. For each resolution pair, there is a joint precipitation amount distribution at high resolution, \(Z_H\), and another at joint low resolution, \(Z_L\), as shown in Fig.~\ref{fig4}a for 0.1º and 0.25º. Here, \(Z_H\) represents how 0.1º and 0.25º rain rates contribute to the precipitation amount at 0.1º, and \(Z_L\) represents how those rain rates contribute to the precipitation amount at 0.25º. \(Z_H\) falls near the one-to-one line indicating that the \(R_H^c\) and \(R_L^c\) contributions to the precipitation amount at 0.1º are closely matched, though they are not identical. Similarly, \(Z_L\) also falls close to the one-to-one line, and mostly overlaps \(Z_H\).

Looking at the broad patterns among the resolution pairs, while \(Z_H\) mostly falls near the one-to-one line for small resolution differences, as the difference in resolution increases, the fraction of the \(Z_H\) falling above the one-to-one line increases. The offset of \(Z_H\) increases because the contributions from \(R_H^c\) and \(R_L^c\) are not identical, with contributions to \(Z_H\) coming from increasingly lighter \(R_L^c\) than \(R_H^c\). The shape of \(Z_H\) also changes for larger differences in resolution as the ranges from which both \(R_H^c\) and \(R_L^c\) contribute decrease. This change is clearest in \(Z_L\) for large resolution differences, e.g. Fig.~\ref{fig4}e, where only a narrow range in \(R_L^c\) contributes to the joint distribution. As resolution differences grow, \(Z_H\) shifts above the one-to-one line, while \(Z_L\) moves below it . As a result, the overlap between \(Z_H\) and \(Z_L\) decreases. Overall, small differences in spatial resolution lead to only slight changes in precipitation intensity, whereas large differences in resolution lead to pronounced differences in intensity. In the next section, we present a quantitative approach to assess how precipitation amount distributions differ between spatial resolutions.

\subsection{The shift mode for changes between spatial resolutions}
\begin{figure*}
 \centerline{\includegraphics[width=5.5in,angle=0]{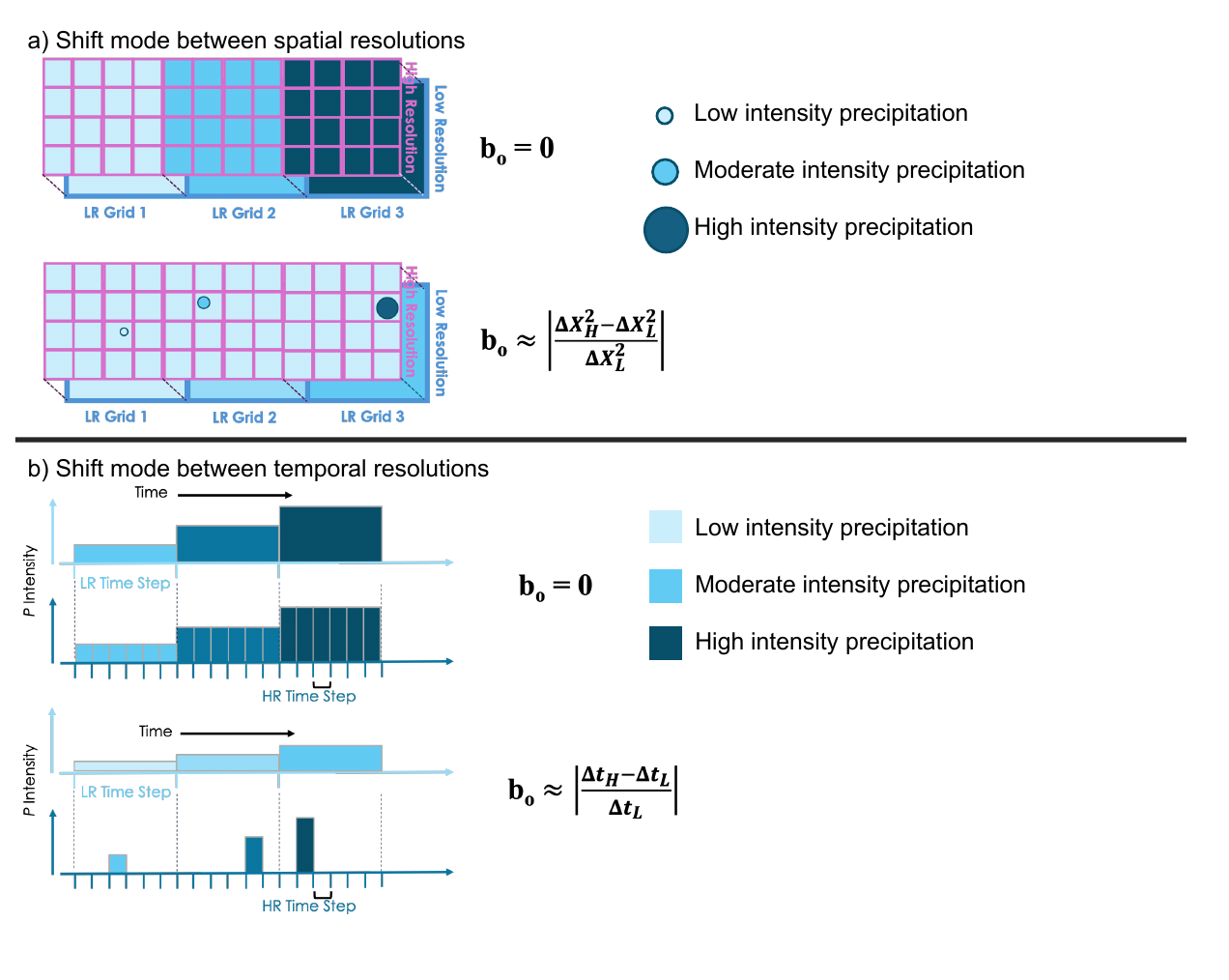}}
\caption{Schematic illustrating different hypothetical cases of the shift mode (\(b_o\)) applied to changes between a) spatial and b) temporal resolutions. Here, high-resolution grids are bordered by pink and low-resolution grids by blue. The circles within individual grid cells represent subgrid-scale precipitation. In both a) and b), darker shades represent higher precipitation intensities at both spatial and temporal resolutions.}\label{fig5}
\end{figure*}

To illustrate the effect of changing resolution, consider two hypothetical cases that bound the possibilities for the shift mode (Fig.~\ref{fig5}a). First, consider the case where a precipitating system is fully resolved at low spatial resolution (Fig.~\ref{fig5}a, top). At higher resolution, there is no change in the distribution; the shift is 0 (\(b_o=0\)). 

Second, consider the case where precipitation events occur at a single point (Fig.~\ref{fig5}a, bottom). For a grid at any given resolution, the pointwise precipitation is spread over the grid cell. At any particular time, some (or many) grid cells may not have any precipitation-generating points, and for simplicity assume there is never more than one point per grid cell. This yields more intense precipitation the smaller the high-resolution grid cell. In this case, the shift mode from the lower to the higher resolution grid would be positive. 
Specifically, it should be
\begin{equation} \label{eq:2}
b_o\approx \Bigg| \frac{\Delta X_H^2-\Delta X_L^2}{\Delta X_L^2}\Bigg|
\end{equation}
where, \(\Delta X_H\) is the length of a high-resolution grid cell, \(\Delta X_H^2\) is its area, and \(\Delta X_L\) is the length of a low-resolution grid cell and \(\Delta X_L^2\) its area.

These bounding cases show that a smaller magnitude of the shift mode indicates that the precipitation distribution is well resolved even on the lower-resolution grid, whereas a larger shift mode indicates larger differences in the precipitation amount distribution between resolutions.

\subsection{Quantifying the shift mode in IMERG}
\begin{figure*}
 \centerline{\includegraphics[width=6.5in,angle=0]{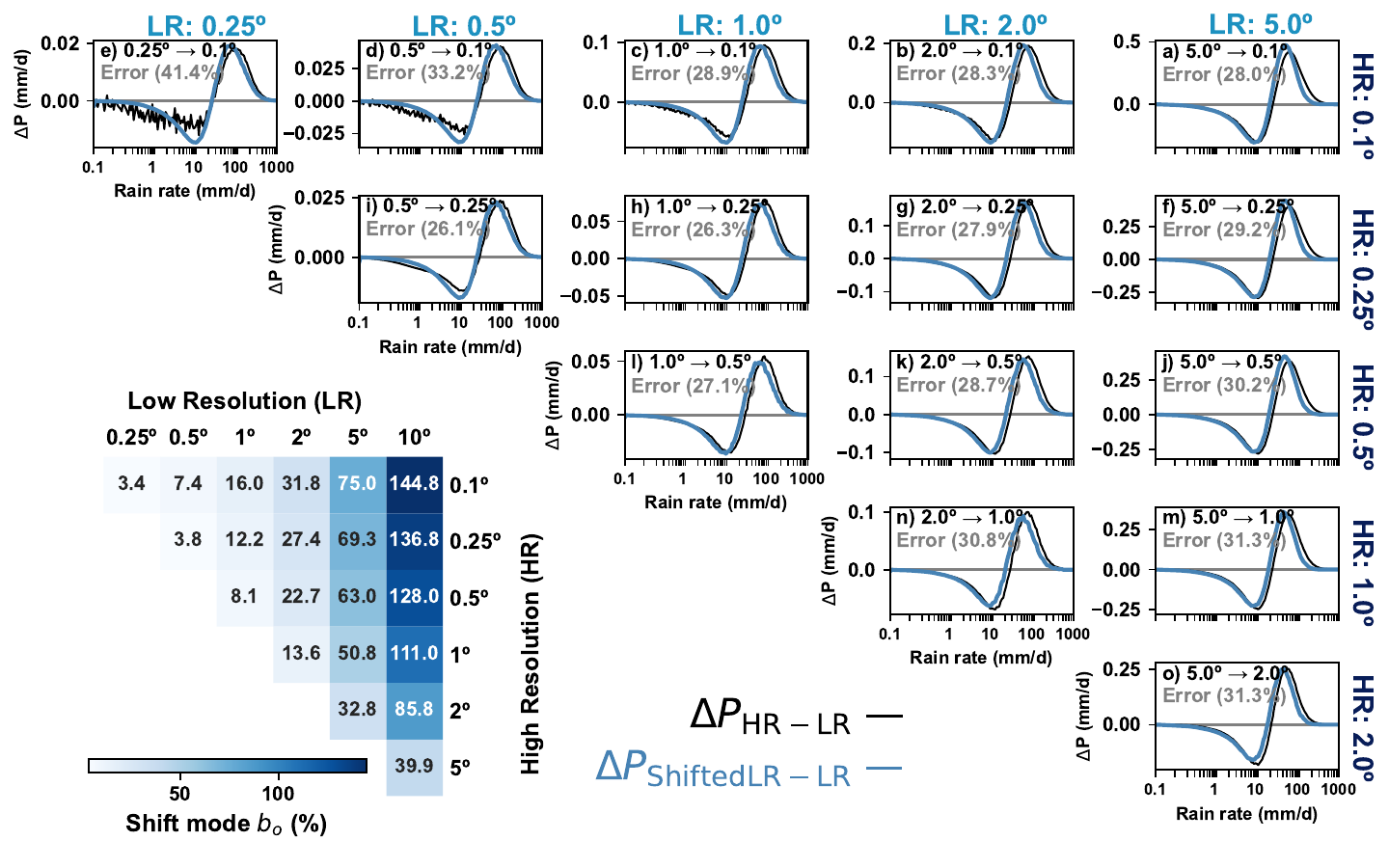}}
\caption{The difference in precipitation amount across rain rates between all high and low-resolution pairs (in black) and the shifted low-resolution and low-resolution (blue). The errors associated with shifted distributions are shown in the top left of each plot. The matrix on the left bottom shows the magnitude of the shift mode (\(b_o\)) for different spatial resolution pairs with darker shades indicating greater shift to heavier rain rates at higher resolutions.}\label{fig6}
\end{figure*}

We apply the shift mode to the spatially aggregated precipitation amount distributions at different grid sizes quantify the effect of spatial resolution difference on precipitation. We first calculate the magnitude of the shift mode, \(b_o\), between two resolutions using Equation~\eqref{eq:1}. Subsequently, we use the magnitude of the shift mode to construct the shifted low-resolution distribution following the method stated in Section~\ref{sec2}e. Finally, we compare the differences between the high-resolution and the shifted low-resolution distributions for each resolution pair, which is shown in Fig.~\ref{fig6}. 

We use an error metric to quantify how well transformation with \(b_o\) captures the differences between resolution pairs. The error metric follows \citet{pendergrass_two_2014}; it is defined as the sum of the absolute differences between \(\Delta p^s\) and \(\Delta p\) at each precipitation rate, normalized by the sum of the absolute values of \(\Delta p\), where \(\Delta p\) is the difference between the native high- and low-resolution precipitation distributions and \(\Delta p^s\) is the difference between the shifted and native low-resolution distributions. This metric describes how much of the precipitation the shifted distribution fails to capture. We calculate the error of the fit for each resolution pair, which is shown as annotation on each panel (a-o) in Fig.~\ref{fig6}. The error for most resolution pairs is between 25\% to 30\%. For the resolution pairs of 0.5º-0.1º and 0.25º-0.1º, the errors are higher, due to noise, which is an artifact of the IMERG V07 algorithm and is removed when regridded to lower resolutions. The shift mode captures most of the difference in precipitation rates between resolution pairs, indicating that the resolution dependent changes in precipitation intensity can be largely characterized by the shift mode. 

To assess how accurately high-resolution precipitation can be recovered using the shift mode to transform from low back to higher resolution, we compare the extreme precipitation from the original high-resolution and shifted distributions (Table~\ref{t1}). Extreme precipitation is estimated from the amount distribution by calculating the precipitation rate that corresponds to 99.7th all-day percentile, which is an approximation of RX1day. Extreme precipitation at 0.1º is only 2.39\% larger than that from the adjusted 0.25º distribution, whereas it is 80.90\% larger than the extreme precipitation from the adjusted 5º distribution. This indicates that the shift mode captures extreme precipitation intensity reasonably well for small resolution differences but not for large differences in resolution.

\begin{table*}
\caption{Comparison of extreme precipitation between precipitation amount distribution from original resolutions against the shifted distributions from IMERG. Extreme precipitation is estimated from the precipitation amount distributions as the precipitation rate that corresponds to 99.7th percentile.}
\label{t1}
\begin{tabular*}{\hsize}{@{\extracolsep{\fill}}ccccc}
\topline
HR & HR Extreme Precipitation (mm/day) & LR & Shifted LR Extreme (mm/day) & Difference (\%) \\
\midline
\multirow{6}{2em}{0.1º} & \multirow{6}{3em}{298.95} & 0.25º & 291.96 & 2.39 \\
    &   & 0.5º & 284.02 & 5.25 \\
     &  & 1º & 266.67 & 12.10 \\
 &  & 2º & 232.83 & 28.40 \\
     &  & 5º & 165.26 & 80.90 \\
     &  & 10º & 119.18  & 150.83 \\
\midline
\multirow{5}{2em}{0.25º} & \multirow{5}{3em}{281.61} & 0.5º & 276.05 & 2.01 \\
 &  & 1º & 258.76 & 8.83 \\
 &  & 2º & 225.15 & 25.08 \\
      &  & 5º & 159.55 & 76.50 \\
      &  & 10º & 115.46  & 143.89 \\
\midline
\multirow{4}{2em}{0.5º} & \multirow{4}{3em}{264.58} & 1º & 248.54 & 6.45 \\
 &  & 2º & 216.86 & 22.00 \\
      &  & 5º & 153.89 & 71.93 \\
      &  & 10º & 111.10  & 138.14 \\
\midline
\multirow{3}{1em}{1º} & \multirow{3}{3em}{229.81} & 2º & 200.69 & 14.51 \\
   &  & 5º & 142.45 & 61.32 \\
      &  & 10º & 102.70  & 123.76 \\
\midline
\multirow{2}{1em}{2º} & \multirow{2}{3em}{176.51} & 5º & 125.06 & 41.14\\ 
&  & 10º & 90.74  & 94.52 \\  
\midline
5º    & 93.98 & 10º  & 68.30  & 37.60 \\
\botline
\end{tabular*}
\end{table*}

Not just extreme precipitation, but the entire distribution of precipitation changes with spatial resolution. To quantitatively encapsulate how the total precipitation distribution changes between two spatial resolutions we focus on the magnitude of the shift mode, shown in Fig.~\ref{fig6}. The shift mode is positive as we move from low to high resolution reflecting precipitation intensification, and its magnitude increases as the difference between grid size increases. For instance, the magnitude of the shift mode going from 0.25º to 0.1º is 3.4\%, indicating that precipitation intensifies by 3.4\%, whereas going from 10º to 0.1º, precipitation intensifies by 144.8\%. If all else were equal,  Equation~\eqref{eq:2} says that changes in resolution of the same factor should have the same \(b_o\). In the case of halving spatial resolution from 2º to 1º, \(b_o\) is 13.6\%, whereas, halving from 1º to 0.5º it is 8.1\%, and for 0.5º to 0.25º it is 3.8\% - each smaller than the last. This occurs because of the difference in how well resolved precipitation is at the base resolution (Section~\ref{sec3}c). Therefore, the magnitude of \(b_o\) depends on both resolution difference and the base resolution. This picture is consistent with that from the joint distributions (Section~\ref{sec3}b), where we found that only large increases in spatial resolution led to pronounced differences in precipitation intensity.  

\subsection{Comparing directly-calculated and fitted shift modes}
\begin{figure}
 \centerline{\includegraphics[width=3.2in]{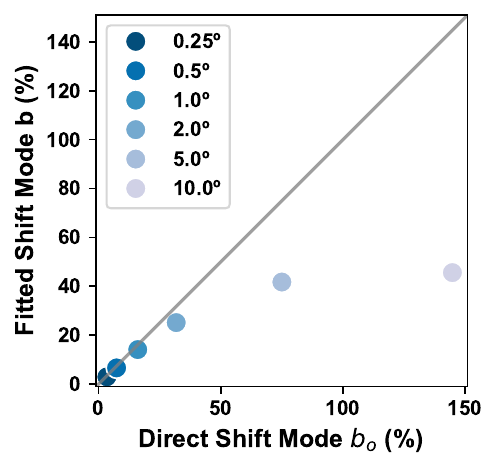}}
\caption{Relationship between fitted (\(b\)) and directly-calculated (\(b_o\)) shift modes. Here the magnitudes of \(b\) and \(b_o\) are calculated between the low-resolution distributions (at 0.25º, 1º, 2º, 5º and 10º) and the precipitation amount distribution at 0.1º.}\label{fig7}
\end{figure}

As mentioned in Section~\ref{sec2}e, our calculation of the shift mode is different from that described in \citet{pendergrass_two_2014}. To investigate the difference between the direct and fitted shift modes, we compare their magnitudes for different spatial resolution pairs (Fig.~\ref{fig7}). The direct and fitted shift modes do not have a linear relationship. For small resolution differences the magnitudes of \(b\) and \(b_o\) are similar. However, as the resolution difference increases, the magnitude of the fitted shift mode saturates at around 50\%, whereas the direct shift mode does not. We interpret this as due to the reliance of the fitting algorithm on the assumption that the differences are small enough to be linear, which breaks down for large resolution differences. In contrast,  \(b_o\) is not limited by such assumptions and continues to work even for large differences in spatial resolution.

\subsection{Quantifying the shift mode in other observational products}
Although we primarily focus our analysis on IMERG, evaluating the consistency of the findings across other observational datasets is essential. To that end, we use a set of multiple merged satellite-gauge datasets from the FROGS precipitation database \citep{roca_frogs_2019}, described in Section~\ref{sec2}a.  We calculate the direct shift mode for different resolution pairs across all datasets in the FROGS database and compare them against IMERG (Table~\ref{t2}). Since the spatial resolution of the datasets is 1º when obtained via FROGS, we compare only against resolutions coarser than 1º. 

\begin{table*}
\caption{Comparison of the magnitude of shift mode (\%) between different spatial resolution pairs for datasets in the FROGS database. The error (\%) of the fit is provide for each resolution pair and product in paranthesis. Here the shift mode (\(b_o\)) is calculated with respect to the lower resolution distribution.}\label{t2}
\begin{tabular*}{\hsize}{@{\extracolsep{\fill}}cclllll}
\topline
HR& LR& IMERG& GPCP& CMORPH& GIRAFE& PERSIANN \\
\midline
1º & 2º & 13.83 (30.1) & 15.47 (27.9)& 16.69 (15.9)	& 12.90 (19.7) & 8.33 (19.2) \\
1º & 5º & 51.67 (30.9) & 57.65 (28.8) & 63.93 (16.5) & 51.16 (6.4) & 31.90 (25.8) \\
1º & 10º & 112.67 (32.5) & 124.84 (30.8) & 138.88 (21.9) & 113.80 (17.4) & 71.27 (32.4) \\
2º & 5º & 33.25 (31.1)	& 36.53 (29.2)	& 40.49 (17.4)	& 33.89 (14.6)	& 21.76 (29.2) \\
2º & 10º & 86.83 (32.3) & 94.73 (30.9) & 104.71 (23.1) & 89.37 (23.0) & 58.11 (34.8) \\
5º & 10º & 40.21 (32.2) & 42.62 (31.1) & 45.72 (26.8) & 41.44 (29.3) & 29.85 (37.7) \\
\botline
\end{tabular*}
\end{table*}

Comparing similar resolution pairs from IMERG and the datasets in FROGS reveals only small differences in the magnitude of \(b_o\) among datasets. These differences may arise from differences in the merging algorithms used to construct the datasets. Different precipitation products have large differences in precipitation properties \citep{bador_diverse_2020, masunaga_interproduct_2019}, so relative agreement is reassuring. Furthermore, the magnitude of \(b_o\) remains proportional to the resolution difference and base resolution, consistent with the behavior observed in IMERG. We also calculate the error metric associated with the shifting of the distributions between resolution pairs for each observational product (Table~\ref{t2}). The error across different products are comparable to or less than that of IMERG, which indicates that the shift mode works comparably well when applied to other observations to quantify differences between spatial resolutions. Overall, the consistency of the shift mode across multiple observational datasets is an indication that these findings reflect properties inherent to actual precipitation rather than characteristics of a particular product.

\subsection{Spatial patterns of the shift mode in IMERG}
\begin{figure*}
 \centerline{\includegraphics[width=5.5in,angle=0]{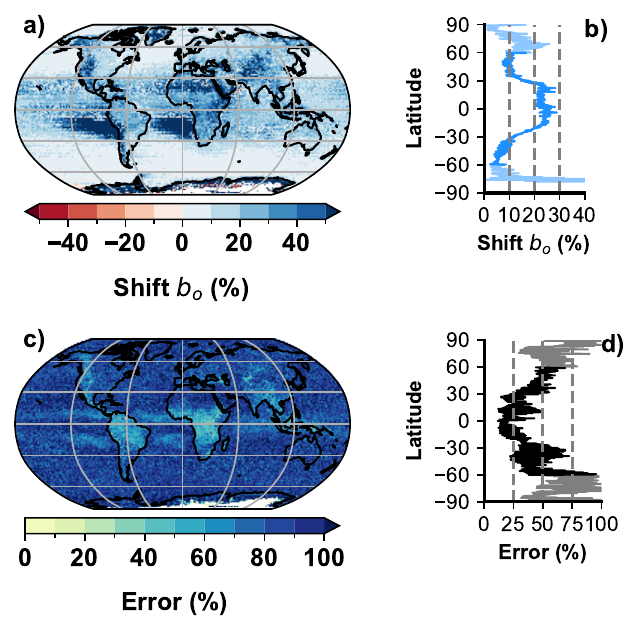}}
\caption{Magnitude of shift mode (\(b_o\)) at each grid point. a) Spatial distribution of the shift mode between spatial resolutions of 1º and 0.1º, b) the shift mode of the zonal-mean precipitation distribution between spatial resolutions of 1º and 0.1º, c) spatial distribution of error of the fit using shift mode between spatial resolutions of 1º and 0.1º and d) the error of the fit using shift mode of zonal-mean precipitation distribution between spatial resolutions of 1º and 0.1º. Positive (negative) shift values correspond to increased (decreased) precipitation rates at high resolution.}\label{fig8}
\end{figure*}

Now that we have applied the shift mode to quantify the changes in precipitation with spatial resolution and tested the results in multiple observational datasets, we return to IMERG to examine the geographic pattern of the shift mode (Fig.~\ref{fig8}a). The spatial pattern of \(b_o\) helps us understand how changes in resolution differs with location. For this analysis, we focus on just the native 0.1º and the coarsened 1º spatial resolutions. To do this, we calculate the precipitation amount distribution at each high-resolution grid point, and match it with the low-resolution grid cell within which it falls. Then, we calculate \(b_o\) and the error of the fit between 1º and 0.1º at each high-resolution grid cell. We also calculate the shift mode of the zonal-mean amount distribution for the same resolution pair (Fig.~\ref{fig8}b).

The shift mode is positive in most places (Fig.~\ref{fig8}a), indicating that the precipitation amount distribution is centred at more intense precipitation rates on the 0.1º grid than the 1º grid. The shift mode of the zonal-mean amount distribution also reflects this and shows that the magnitude of \(b_o\) is larger over the tropics. In contrast, the shift mode with increasing resolution is generally lower in magnitude in the extratropics. Similar to the zonal contrast, Fig.~\ref{fig8}a shows that \(b_o\) is larger over land than over ocean. Furthermore, over land regions, grid cells associated with high orography also have a large \(b_o\). We also note that stratocumulus cloud decks have some consistently large values of \(b_o\), however, IMERG should be treated with caution in these regions of mostly very light precipitation.

Prior to interpreting what might cause \(b_o\) to vary spatially, we assess how well \(b_o\) captures the difference between resolution pairs using the error metric introduced in Section~\ref{sec3}d. We calculate the error of the fit of the shift mode at each grid point using nearest neighbor interpolation on the coarser resolution so that both resolutions overlap perfectly (Fig.~\ref{fig8}c). The errors are lower over the tropics, especially over land, and increase in the extratropics. The errors are large at most grid cells because precipitation amount distributions at individual grid cells are noisy. Zonally averaging helps smooth out these distributions further, yielding relatively lower error values, as shown in Fig.~\ref{fig8}d. The relatively short length of the dataset contributes to the noise; as the record continues to lengthen going forward, this could further reduce noise at the grid level and improve the fit by producing smoother distributions.

The spatial variations in the magnitude of \(b_o\) indicate that increasing spatial resolution does not affect all regions equally. As discussed in Section~\ref{sec3}c, a smaller magnitude of \(b_o\) indicates that precipitating systems are already relatively well-resolved at the lower resolution. The spatial pattern in the magnitude of \(b_o\) therefore reflects how well precipitation at 0.1º is resolved on a 1º grid across different regions. The zonal variation in the magnitude of \(b_o\) indicates that both 1º and 0.1º grids resolve precipitation in the extratropics, but not in the tropics. Similarly, the land-ocean contrast highlights how 1º and 0.1º grids resolves precipitation over land and ocean differently, reflecting the difference in spatial scale of precipitating systems there.

\section{Dependence of precipitation intensity on temporal resolution}
\label{sec4}

Just as we investigated the effect of spatial resolution on precipitation intensity, we explore how precipitation changes over different accumulation periods in time. Here, we pivot from merged satellite-gauge datasets to DOE ARM gauge data with a native temporal resolution of 1-minute and use it to calculate precipitation amount distribution at different temporal resolutions, described in Section~\ref{sec2}c. The precipitation amount distributions at native and various coarsened temporal resolutions over the TWP and SGP locations are shown in Fig.~\ref{fig3}. We note that the distributions have substantial noise which diminishes with coarser temporal resolution, as more time steps are included within the averaging window. Focusing on the precipitation amount peak (\(r_{1/2}\); defined in Section~\ref{sec3}), at TWP, \(r_{1/2}\) is at 0.03 mm/min (equivalent to 43 mm/day) for the 24-hour distribution but for the 1-minute distribution it moves to 0.62 mm/min. For SGP, \(r_{1/2}\) of the 24-hour distribution is at 0.02 mm/min (or 28 mm/day) and moves to 0.52 mm/min for the 1-minute distribution. The increase in the intensity at which \(r_{1/2}\) occurs with temporal resolution reflects the shift of the entire distribution to heavier rain rates as resolution increases.

\subsection{Application of joint precipitation amount distribution to changing temporal resolution}

\begin{figure*}
 \centerline{\includegraphics[width=6.5in,angle=0]{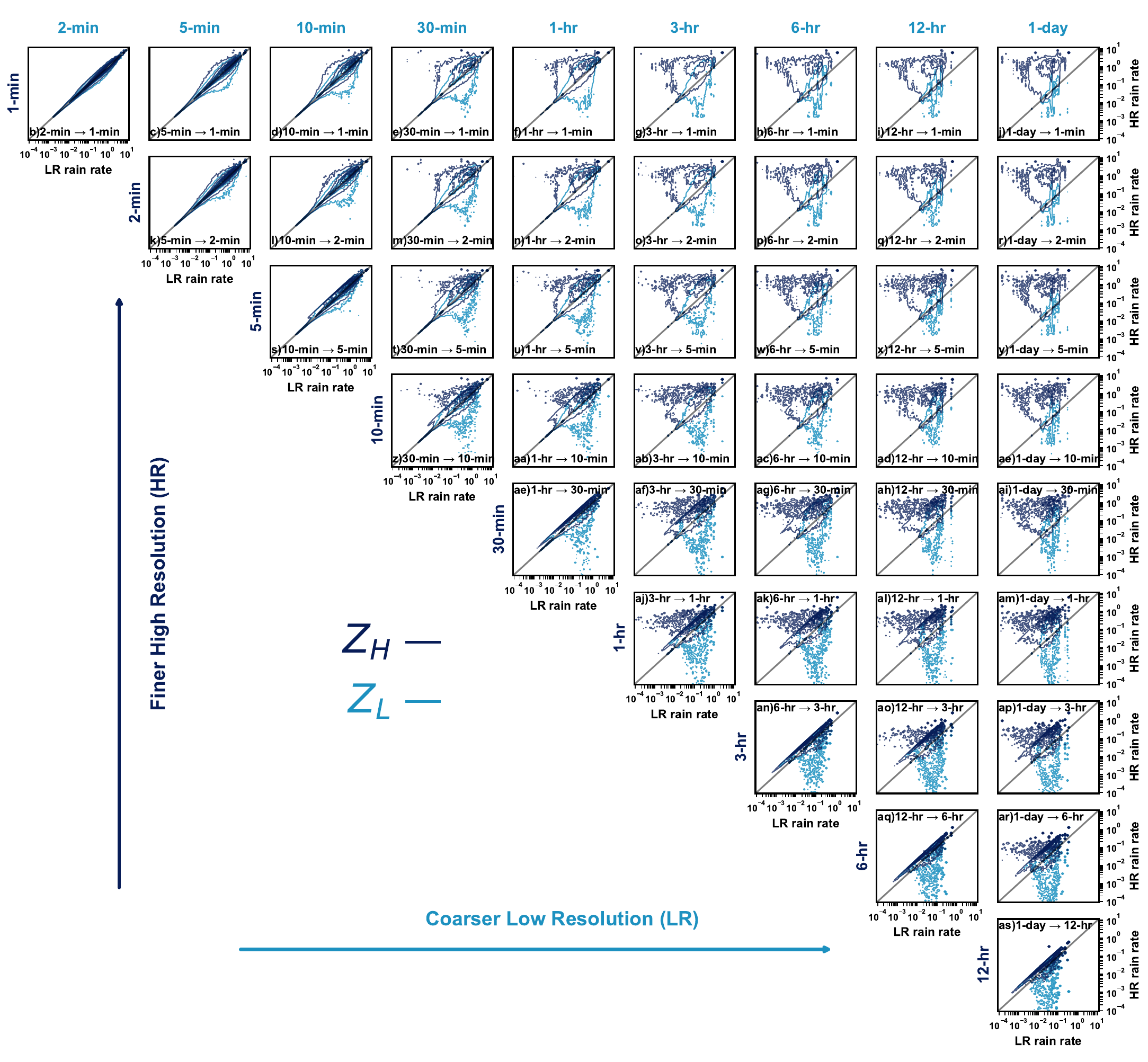}}
\caption{Joint precipitation amount distributions between high and low temporal resolutions from DOE ARM gauge data at Manus Island (TWP), coarsened to different resolutions. The high-resolution datasets used are 1-minute, 2-minute, 5-minute, 10-minute, 30-minute, 1-hour, 3-hour, 6-hour and 12-hour. Each row corresponds to a fixed high-resolution, and each column corresponds to a fixed low-resolution. Rain rate bins from the low-resolution data are plotted along the $x$-axis, and those from the high-resolution data are along the $y$-axis. Here, dark blue contours indicate total precipitation at the joint occurrence of high and low-resolution rain rate bins evaluated over the high-resolution precipitation amount \(Z_H\), while light blue contours indicate the same evaluated over the low-resolution precipitation amount \(Z_L\).}\label{fig9}
\end{figure*}

To further understand how precipitation amount changes with temporal resolution, we construct the joint precipitation amount distribution between high and low temporal resolutions. In order to accomplish this, we first match each high-resolution time step with the low-resolution time step in which it is contained. We then calculate the joint distributions for different temporal resolution pairs following the description in Section~\ref{sec2}d, shown in Fig.~\ref{fig9}. For the resolution pair of 1-minute and 2-minute, the high-resolution joint precipitation amount distribution (\(Z_H\)) represents how 1- and 2-minute resolution precipitation rates (\(R_H^c\) \& \(R_L^c\)) contribute to the total precipitation at 1-minute temporal resolution. Similarly, the low-resolution joint precipitation amount distribution (\(Z_L\)) represents how 1- and 2-minute resolution precipitation rates contribute to the total precipitation at 2-minute temporal resolution. In the case of this resolution pair, \(Z_H\) falls along the one-to-one line signifying that both \(R_H^c\) and \(R_L^c\) contributes nearly equally. \(Z_L\) mostly overlaps with \(Z_H\). 

As the difference between the resolution pairs increases, such as between 5-minute and 12-hour temporal resolutions, both \(Z_H\) and \(Z_L\) move away from the one-to-one line signifying larger differences in the contribution of \(R_H^c\) and \(R_L^c\) to the total precipitation between the temporal resolutions. \(Z_H\) and \(Z_L\) change between temporal resolutions in a similar manner as they did between spatial resolutions. As the resolution difference increases, \(Z_H\) shifts above the one-to-one line, whereas \(Z_L\) falls below it, while the contribution of \(R_L^c\) to \(Z_L\) becomes limited to a narrow range. However, when temporal resolution doubles, for instance 6- to 3- hours, 12- to 6-hours and 1-day to 12-hours, \(Z_H\) is slightly offset from the one-to-one line but falls parallel to it, signifying that the contribution of both \(R_H^c\) and \(R_L^c\) do not change substantially. This offset decreases when the base resolution is higher, such as from 10- to 5-minute and from 2- to 1-minute resolutions. This indicates that \(Z_H\) does not change substantially when temporal resolution is increased by a factor of 2. However, doubling the temporal resolution does not have a similar impact on \(Z_L\). For instance, the extent by which \(Z_L\) in the 10- to 5-minute and 1-hour to 30-minute resolution pairs move away from the one-to-one line seem to reflect the resolution difference, while \(Z_H\) for the same resolution pairs do not differ drastically. While the changes in \(Z_H\) and \(Z_L\) for different resolution pairs show us that precipitation intensity does depend on temporal resolution, the similarity in \(Z_H\) across certain resolution pairs indicates that precipitation is governed by the ratio between resolutions in addition to the absolute difference. The changes in \(Z_H\) and \(Z_L\) between different temporal resolutions are reflected in the total precipitation amount distributions. In the next sub-section, we use the total precipitation amount distribution to quantitatively assess how precipitation intensity differs between temporal resolutions.

\subsection{The shift mode for changes between temporal resolutions: Theory}
Analogously to the effect of changing spatial resolution, we can apply the shift mode to quantify the differences in precipitation amount distributions with temporal resolutions (Fig.~\ref{fig5}b). To illustrate how we interpret this, first consider two hypothetical bounding cases. In the first case, precipitation is distributed equally within each lower temporal resolution timestep and over all finer time steps, the high- and low-resolution precipitation amount distributions will be similar. Then the shift mode is 0 (Fig.~\ref{fig5}b, top). On the other hand, if all of the precipitation occurs at a single point in time, within one high-temporal-resolution time step, and no precipitation falls in the other time steps, then the shift mode for the difference from the lower to the higher resolution grid should be the relative difference between high and low temporal resolutions. Specifically, 
\begin{equation} \label{eq:3}
b_o\approx\Bigg| \frac{\Delta t_H-\Delta t_L}{\Delta t_L}\Bigg|
\end{equation}
where, \(\Delta t_H\) is the timestep for high temporal resolution and \(\Delta t_L\) is the timestep for low temporal resolution. Like we saw for spatial resolution, a smaller magnitude of \(b_o\) indicates that the precipitation intensity distribution is similar at both high and low resolutions, while a larger magnitude of \(b_o\) represents that it differs more between two temporal resolutions. However, the form of the scaling differs from the spatial case; it is the difference in \(\Delta t\) rather than \(\Delta X^2\).

\subsection{Application of the shift mode to changing temporal resolution}

\begin{figure*}
 \centerline{\includegraphics[width=6.5in,angle=0]{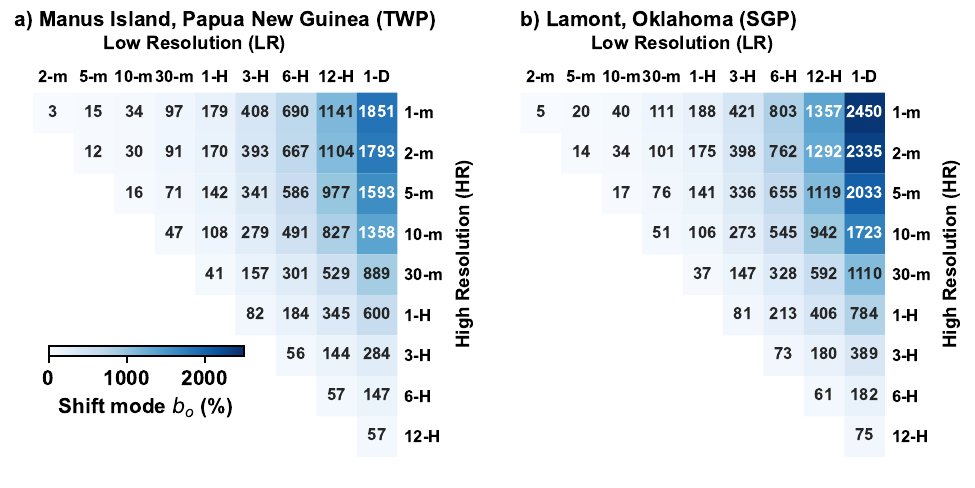}}
\caption{Magnitude of the shift mode between of the precipitation amount distributions at different temporal resolutions over two locations. Matrix of the magnitudes of the directly-calculated shift mode (\(b_o\) for different temporal resolution pairs for a) Manus Island (TWP), and for b) Lamont, Oklahoma (SGP). Here darker shades indicating greater shift to heavier rain rates at higher resolutions.}\label{fig10}
\end{figure*}

We again use the directly-calculated shift mode, \(b_o\), to quantify the changes between precipitation amount distribution at different temporal resolutions, as described in Section~\ref{sec4}b. The magnitude of the shift mode for a range of temporal-resolution pairs at TWP and SGP is shown in Fig.~\ref{fig10}. Analogously to spatial resolution, the shift mode is positive when moving from low to high temporal resolution. At TWP, the shift in the precipitation amount distribution from 2-minute to 1-minute is 3\% and increase to 1851\% when going from 1-day to 1-minute temporal resolution. At SGP, the magnitude of \(b_o\) for the same resolution pairs are 5\% and 2450\%. Moreover, when temporal resolution is doubled at coarser absolute resolutions, the magnitude of the shift mode is only marginally different. For instance, at TWP, the magnitude of \(b_o\) is 57\% for each of the 1-day to 12-hr, 12- to 6-hr, and 6- to 3-hr doublings, whereas at SGP the magnitudes for the same three pairs are 75\%, 61\%, and 73\%, respectively. However, at higher temporal resolutions, such as doubling from 10- to 5-minute and from 2- to 1-minute, the magnitude of \(b_o\) is drastically different with values of 16\% and 3\% at TWP, and 17\% and 5\% at SGP, respectively. We also note that the magnitudes of \(b_o\) for corresponding resolution pairs are larger at SGP compared to those at TWP, which could be due to the difference in the length of the datasets or other factors. Overall, the behavior of precipitation intensity across temporal and spatial resolution differences are analogous in that they depend on both the change in resolution and the base resolution.

\section{Discussion}
\label{sec5}

Characterizing the effect of the changes between spatial and temporal resolution on the distribution of precipitation is a long-standing challenge with implications for our understanding of atmospheric processes. The results presented in this study addresses this challenge by showing that the precipitation amount distribution shifts to heavier precipitation rates at higher spatial and temporal resolutions, with little change in shape. Previous studies, particularly those comparing precipitation in climate models at increased resolution have largely attributed such changes in precipitation intensity to improvements in the representation of physical processes at higher resolution \citep[e.g.,][]{kopparla_improved_2013,slingo_ambitious_2022,jong_increases_2023,lin_resolution_2023,nikumbh_does_2024}. We demonstrate that these changes are an intrinsic property of resolution rather than an improvement in modeled physics, as our analysis compare within datasets by changing only their resolutions, not the underyling physics (which is held fixed by construction). Furthermore, we show that the behavior can be characterized to zeroth or possibly first order using the shift mode.

The reduction in the intensity of the precipitation amount peak (\(r_{1/2}\)) at lower spatial and temporal resolutions occurs because coarsening averages over intense precipitation at smaller grids or time steps \citep{rajulapati_perils_2021}. Moreover, the magnitude of the shift mode does not depend solely on the base resolution and its differences, but also on the spatial and temporal structure of precipitation intensity. This is evident in the zonal contrast of the shift mode (\(b_o\)) for spatial resolution-dependent changes (Section~\ref{sec3}g), as the coarser resolution does not capture the smaller scale, more convective precipitating systems in the tropics \citep{dai_precipitation_2006,schumacher_stratiform_2003}, yielding larger magnitudes of the shift mode, while large-scale stratiform precipitation dominates in the extratropics, which is resolved at both resolutions, so the shift mode remains comparatively small. A similar contrast of the shift mode for spatial resolution-dependent differences exists between the land and the ocean. Over land, continental convection dominates \citep{xu_properties_2012}, characterized by smaller scale processes that would not be resolved at coarser resolutions; whereas over the ocean, precipitating systems have larger spatial footprints and thus better resolved at both high and low resolutions. 

The temporal resolution analysis shows that the magnitude of the shift mode per doubling is not constant across the full range of temporal resolutions. This behavior is reflected at two stations and is consistent with the multi-scaling behavior discussed in \citet{deidda_rainfall_2000} and \citet{veneziano_imperfect_2006}. The differences in the magnitude of the shift mode between temporal resolutions are consistent with the findings of \citet{martinez-villalobos_why_2019}, who demonstrated that the shape of the precipitation intensity distribution can be viewed as controlled by the frequency and total accumulation of precipitation events, both of which vary with temporal resolution.

While we show that the shift mode can be used quantify the differences between spatial and temporal resolutions, it is far from perfect. For example, it does not fully recover heavy precipitation rates, especially over large resolution differences (Section~\ref{sec3}d, Table~\ref{t1}). This limitation may stem from the fact that the directly-calculated shift mode \(b_o\), is based on the precipitation rate at which the bulk of precipitation occurs, whereas extreme precipitation may not change at the same rate as moderate rain rates. Such scale-dependent behavior is consistent with previous findings suggesting that precipitation exhibits multi-fractal characteristics, wherein different intensity ranges scale inconsistently across resolutions \citep{marani_nonpowerlawscale_2005,veneziano_imperfect_2006}. Nevertheless, \(b_o\) provides a simple framework to quantify changes between precipitation amount distributions across spatial or temporal resolution.

One caveat worth mentioning is that our findings associated with changes between spatial resolutions are based on merged-gauge observational datasets and would be influenced by the effective spatial resolution of the product \citep{guilloteau_multiscale_2020}. Furthermore, it is possible that precipitation intensity changes at spatial resolutions finer than 0.1º may differ substantially from those reported here, but cannot be assessed given the resolution limitations of currently available precipitation products. 

In the case of the temporal resolution analysis, it should be noted that the TWP site uses optical rain gauges which have been shown to exhibit a positive bias, tending to overestimate precipitation during periods of extreme precipitation \citep{nystuen_relative_1999}. The SGP location uses a weighing rain gauge and avoid biases at extreme rates. However, bias in their estimates from wind or turbulence induced undercatch and wetting losses, which lead to underestimation in many real-world operational conditions \citep{colli_performance_2013}. 

The precipitation-amount-distribution-based framework introduced in this study can be leveraged to evaluate precipitation simulations at higher resolutions, rather than coarsening them to a common low-resolution grid \citep{bador_impact_2020}, and could provide a baseline for model evaluation that would reveal the added value of higher resolution. This is important because the sensitivity of the precipitation intensity distribution to grid size indicates that comparisons across products with differing spatial resolutions may yield misleading assessments of dataset performance if resolution dependence is not accounted for. Finally, this observational analysis serves as a baseline for future investigations on how precipitation intensity responds to changes in spatial and temporal resolution in both models and other observational products.

\section{Conclusions}\label{sec6}
This study aims to understand and quantify the changes in the distribution of precipitation intensity as spatial and temporal resolution changes in observational products. Much of the changes in the precipitation amount distribution between two resolutions can be characterized quite well by applying the shift mode to the precipitation amount distribution, without changing its shape. We apply the shift mode to quantify the difference between spatial and temporal resolutions, and its magnitude reflects the relative intensification of precipitation between two resolutions. The magnitude of the shift mode depends on the base resolution, the magnitude of resolution difference and the structure of precipitation intensity in space or time, with regional differences that reflect the dominant spatial scale of precipitating systems in the case of spatial resolution differences. These findings challenge the notion that precipitation intensification at higher resolutions solely reflects improvements in the representation of fine-scale processes, demonstrating instead that intensity changes with resolution even in the absence of changes in the underlying physics.

The framework introduced here could be used to isolate the effects of resolution, and perform model evaluation at the native high resolution, where the added value of increased resolution should be most apparent, particularly given the recent development of kilometer-scale climate models. Similarly, this framework has the potential to evolve into a downscaling tool with additional research, allowing existing low-resolution datasets to provide higher resolution information without the computational expense of high-resolution simulations. This study provides an observational baseline against which the resolution-dependent behavior of precipitation datasets can be evaluated. Future work should also extend this framework to spatial resolutions higher than 0.1º, where convective processes are explicitly resolved, when such observational datasets become available. Finally, this framework could potentially be used to develop a generalized spatial and temporal scaling relationship for precipitation intensity.

\clearpage
\acknowledgments
This work was funded by the U.S. National Science Foundation under Grant number AGS-2441755. This work was supported in part by a Sloan Research Fellowship from the Alfred P. Sloan Foundation. The analysis code used in this study is archived on Zenodo at https://doi.org/10.5281/zenodo.22905614.

%
%
\datastatement
All datasets used in this work are publicly available. The IMERG daily precipitation data (Final Run version) used in this study are publicly available from the NASA Goddard Earth Sciences Data and Information Services Center (GES DISC) at \url{https://gpm.nasa.gov/data/directory}.
The FROGS (Frequent Rainfall Observations on GridS) database is publicly accessible and maintained by IRD/IRD-LTHE, and can be downloaded from \url{https://frogs.ird.fr}. The data from the in-situ observations are available from the US Department of Energy's Atmospheric Radiation Measurement (ARM) program at \url{https://armgov.svcs.arm.gov/data}. The analysis code and mid-level processed data required to reproduce the figures in this study are available on Github at \url{https://github.com/akshayrajeev1/precip-resolution-obs}.


\appendix
\appendixtitle{Calculation of the joint precipitation amount distribution}

We calculate the joint precipitation amount distribution \(Z_H(R_H^c,R_L^c)\) and \(Z_L(R_H^c,R_L^c)\) using daily precipitation accumulations at high- and low-spatial resolution datasets. \(R^l\) and \(R^r\) are the left and right bin edges, and \(R^c = (R^r+R^l)/2\) is the bin center,  which we use in transforming the gridded dataset into the distributions. The bins are constructed for both high and low resolutions which we indicate using the subscripts \(H\) and \(L\). The high-spatial-resolution joint distribution is,

\begin{equation} \label{eq:A1}
Z_H(R_H^c,R_L^c)=
\frac{1}{N_LN_T \left(\frac{\Delta X_L}{\Delta X_H}\right)^2 }\sum_{i}\sum_{j}\sum_{k} r_{H(i,k)} (R_H^l< r_{i,k} < R_H^r | R_L^l< r_{j,k} < R_L^r)
\end{equation}

and the low-spatial-resolution joint distribution is,

\begin{equation} \label{eq:A2}
Z_L(R_H^c,R_L^c) =
\frac{1}{N_LN_T \left(\frac{\Delta X_L}{\Delta X_H}\right)^2} \sum_{i}\sum_{j}\sum_{k} r_{L(j,k)} (R_H^l< r_{i,k} < R_H^r | R_L^l< r_{j,k} < R_L^r)\\
\end{equation}

Here \(r_{H(i,k)}\) is the precipitation accumulation at the \(i^{th}\) grid point on the high-resolution grid and  \(r_{L(j,k)}\) is the precipitation accumulation at the \(j^{th}\) grid point on the low-resolution grids, each at the \(k^{th}\) timestep. \(\Delta X_H\) and \(\Delta X_L\) are the grid sizes at high and low resolution, and the joint precipitation amount distributions calculated at each grid are averaged over all low-resolution grid-cells, of which there are \(N_L\), and normalized by the total number of days \(N_T\).

Similarly, we calculate the joint precipitation amount distributions \(Z_H(R_H^c,R_L^c)\) and \(Z_L(R_H^c,R_L^c)\) using precipitation accumulations at high- and low-temporal resolution datasets. As mentioned in Section~\ref{sec4}a, the number of timesteps in the high- and low-temporal resolution timeseries are regridded so that each high-resolution time step match with the low-resolution time step in which it is contained. The high-temporal-resolution joint distribution is,
\begin{equation} \label{eq:A3}
Z_H(R_H^c,R_L^c)=
\frac{1}{N_T\left(\frac{\Delta T_L}{\Delta T_H}\right)}\sum_{m}\sum_{n} r_{H(m)} (R_H^l< r_m < R_H^r | R_L^l< r_n < R_L^r)
\end{equation}
and the low-temporal-resolution distribution is,
\begin{equation} \label{eq:A4}
Z_L(R_H^c,R_L^c) =
\frac{1}{N_T\left(\frac{\Delta T_L}{\Delta T_H}\right)}\sum_{m}\sum_{n} r_{L(n)} (R_H^l< r_{m} < R_H^r | R_L^l< r_{n} < R_L^r)\\
\end{equation}
Here \(r_{H(m)}\) is the precipitation accumulation at the \(m^{th}\) timestep on the high-resolution timeseries and \(r_{L(n)}\) is the precipitation accumulation at the \(n^{th}\) timestep on the low-resolution timeseries. \(\Delta T_H\) and \(\Delta T_L\) are the lengths of the high and low resolution time steps, and the joint precipitation amount distributions calculated at each time step are averaged over all low-resolution time steps, of which there are \(N_T\).







%



\bibliographystyle{ametsocV6}
\bibliography{references}

@article{alexander_global_2006,
  title = {Global Observed Changes in Daily Climate Extremes of Temperature and Precipitation},
  author = {Alexander, L. V. and Zhang, X. and Peterson, T. C. and Caesar, J. and Gleason, B. and Klein Tank, A. M. G. and Haylock, M. and Collins, D. and Trewin, B. and Rahimzadeh, F. and Tagipour, A. and Rupa Kumar, K. and Revadekar, J. and Griffiths, G. and Vincent, L. and Stephenson, D. B. and Burn, J. and Aguilar, E. and Brunet, M. and Taylor, M. and New, M. and Zhai, P. and Rusticucci, M. and {Vazquez-Aguirre}, J. L.},
  year = 2006,
  journal = {Journal of Geophysical Research: Atmospheres},
  volume = {111},
  number = {D5},
  issn = {2156-2202},
  doi = {10.1029/2005JD006290},
  copyright = {Copyright 2006 by the American Geophysical Union.},
  language = {en}
}

@article{alexander_use_2019,
  title = {On the Use of Indices to Study Extreme Precipitation on Sub-Daily and Daily Timescales},
  author = {Alexander, Lisa V. and Fowler, Hayley J. and Bador, Margot and Behrangi, Ali and Donat, Markus G. and Dunn, Robert and Funk, Chris and Goldie, James and Lewis, Elizabeth and Rog{\'e}, Marine and Seneviratne, Sonia I. and Venugopal, V.},
  year = 2019,
  month = dec,
  journal = {Environmental Research Letters},
  volume = {14},
  number = {12},
  pages = {125008},
  issn = {1748-9326},
  doi = {10.1088/1748-9326/ab51b6},
  language = {en}
}

@article{ashouri_persianncdr_2015,
  title = {{{PERSIANN-CDR}}: {{Daily Precipitation Climate Data Record}} from {{Multisatellite Observations}} for {{Hydrological}} and {{Climate Studies}}},
  shorttitle = {{{PERSIANN-CDR}}},
  author = {Ashouri, Hamed and Hsu, Kuo-Lin and Sorooshian, Soroosh and Braithwaite, Dan K. and Knapp, Kenneth R. and Cecil, L. Dewayne and Nelson, Brian R. and Prat, Olivier P.},
  year = 2015,
  month = jan,
  journal = {Bulletin of the American Meteorological Society},
  volume = {96},
  number = {1},
  pages = {69--83},
  issn = {0003-0007, 1520-0477},
  doi = {10.1175/BAMS-D-13-00068.1},
  chapter = {Bulletin of the American Meteorological Society},
  language = {EN}
}

@article{bador_diverse_2020,
  title = {Diverse Estimates of Annual Maxima Daily Precipitation in 22 State-of-the-Art Quasi-Global Land Observation Datasets},
  author = {Bador, Margot and Alexander, Lisa V and Contractor, Steefan and Roca, Remy},
  year = 2020,
  month = feb,
  journal = {Environmental Research Letters},
  volume = {15},
  number = {3},
  pages = {035005},
  issn = {1748-9326},
  doi = {10.1088/1748-9326/ab6a22},
  language = {en}
}

@article{bador_impact_2020,
  title = {Impact of {{Higher Spatial Atmospheric Resolution}} on {{Precipitation Extremes Over Land}} in {{Global Climate Models}}},
  author = {Bador, Margot and Bo{\'e}, Julien and Terray, Laurent and Alexander, Lisa V. and Baker, Alexander and Bellucci, Alessio and Haarsma, Rein and Koenigk, Torben and Moine, Marie-Pierre and Lohmann, Katja and Putrasahan, Dian A. and Roberts, Chris and Roberts, Malcolm and Scoccimarro, Enrico and Schiemann, Reinhard and Seddon, Jon and Senan, Retish and Valcke, Sophie and Vanniere, Benoit},
  year = 2020,
  month = may,
  journal = {Journal of Geophysical Research: Atmospheres},
  volume = {125},
  number = {13},
  pages = {e2019JD032184},
  issn = {2169-8996},
  doi = {10.1029/2019JD032184},
  language = {en}
}

@article{chen_dependence_2018,
  title = {Dependence of Estimated Precipitation Frequency and Intensity on Data Resolution},
  author = {Chen, Di and Dai, Aiguo},
  year = 2018,
  month = may,
  journal = {Climate Dynamics},
  volume = {50},
  number = {9},
  pages = {3625--3647},
  issn = {1432-0894},
  doi = {10.1007/s00382-017-3830-7},
  language = {en}
}

@article{chen_precipitation_2019,
  title = {Precipitation {{Characteristics}} in the {{Community Atmosphere Model}} and {{Their Dependence}} on {{Model Physics}} and {{Resolution}}},
  author = {Chen, Di and Dai, Aiguo},
  year = 2019,
  journal = {Journal of Advances in Modeling Earth Systems},
  volume = {11},
  number = {7},
  pages = {2352--2374},
  issn = {1942-2466},
  doi = {10.1029/2018MS001536},
  copyright = {\copyright 2019. The Authors.},
  language = {en}
}

@article{colli_performance_2013,
  title = {Performance of a Weighing Rain Gauge under Laboratory Simulated Time-Varying Reference Rainfall Rates},
  author = {Colli, M. and Lanza, L. G. and La Barbera, P.},
  year = 2013,
  month = sep,
  journal = {Atmospheric Research},
  series = {Perspectives of {{Precipitation Science}} - {{Part I}}},
  volume = {131},
  pages = {3--12},
  issn = {0169-8095},
  doi = {10.1016/j.atmosres.2013.04.006}
}

@article{dai_precipitation_2006,
  title = {Precipitation {{Characteristics}} in {{Eighteen Coupled Climate Models}}},
  author = {Dai, Aiguo},
  year = 2006,
  month = sep,
  journal = {Journal of Climate},
  volume = {19},
  number = {18},
  pages = {4605--4630},
  issn = {0894-8755, 1520-0442},
  doi = {10.1175/JCLI3884.1},
  chapter = {Journal of Climate},
  language = {EN}
}

@article{deidda_rainfall_2000,
  title = {Rainfall Downscaling in a Space-Time Multifractal Framework},
  author = {Deidda, R.},
  year = 2000,
  journal = {Water Resources Research},
  volume = {36},
  number = {7},
  pages = {1779--1794},
  issn = {1944-7973},
  doi = {10.1029/2000WR900038},
  language = {en}
}

@article{dewitt_climatologically_2024,
  title = {Climatologically Invariant Scale Invariance Seen in Distributions of Cloud Horizontal Sizes},
  author = {DeWitt, Thomas D. and Garrett, Timothy J. and Rees, Karlie N. and Bois, Corey and Krueger, Steven K. and Ferlay, Nicolas},
  year = 2024,
  month = jan,
  journal = {Atmospheric Chemistry and Physics},
  volume = {24},
  number = {1},
  pages = {109--122},
  issn = {1680-7316},
  doi = {10.5194/acp-24-109-2024},
  language = {English}
}

@article{eggert_temporal_2015,
  title = {Temporal and Spatial Scaling Impacts on Extreme Precipitation},
  author = {Eggert, B. and Berg, P. and Haerter, J. O. and Jacob, D. and Moseley, C.},
  year = 2015,
  month = may,
  journal = {Atmospheric Chemistry and Physics},
  volume = {15},
  number = {10},
  pages = {5957--5971},
  issn = {1680-7316},
  doi = {10.5194/acp-15-5957-2015},
  language = {English}
}

@article{fraedrich_scaling_1993,
  title = {Scaling Regimes of Composite Rainfall Time Series},
  author = {Fraedrich, Klaus and Larnder, Chris},
  year = 1993,
  journal = {Tellus A},
  volume = {45},
  number = {4},
  pages = {289--298},
  issn = {1600-0870},
  doi = {10.1034/j.1600-0870.1993.t01-3-00004.x},
  language = {en}
}

@misc{g.j.huffman_gpm_2023,
  title = {{{GPM IMERG Final Precipitation L3}} 1 Day 0.1 Degree x 0.1 Degree {{V07}}, {{Edited}} by {{Andrey Savtchenko}}, {{Greenbelt}}, {{MD}}, {{Goddard Earth Sciences Data}} and {{Information Services Center}} ({{GES DISC}})},
  author = {{G.J. Huffman} and Stocker, E.F. and Bolvin, D.T. and Nelkin, E.J. and Tan, Jackson},
  year = 2023,
  doi = {10.5067/GPM/IMERGDF/DAY/07}
}

@article{gebremichael_scaling_2008,
  title = {Scaling of Tropical Rainfall as Observed by {{TRMM}} Precipitation Radar},
  author = {Gebremichael, Mekonnen and Krajewski, Witold F. and Over, Thomas M. and Takayabu, Yukari N. and Arkin, Phillip and Katayama, M.},
  year = 2008,
  month = jun,
  journal = {Atmospheric Research},
  volume = {88},
  number = {3},
  pages = {337--354},
  issn = {0169-8095},
  doi = {10.1016/j.atmosres.2007.11.028}
}

@incollection{guilloteau_multiscale_2020,
  title = {Multiscale {{Evaluation}} of {{Satellite Precipitation Products}}: {{Effective Resolution}} of {{IMERG}}},
  shorttitle = {Multiscale {{Evaluation}} of {{Satellite Precipitation Products}}},
  booktitle = {Satellite {{Precipitation Measurement}}: {{Volume}} 2},
  author = {Guilloteau, Cl{\'e}ment and {Foufoula-Georgiou}, Efi},
  editor = {Levizzani, Vincenzo and Kidd, Christopher and Kirschbaum, Dalia B. and Kummerow, Christian D. and Nakamura, Kenji and Turk, F. Joseph},
  year = 2020,
  pages = {533--558},
  address = {Cham},
  doi = {10.1007/978-3-030-35798-6_5},
  isbn = {978-3-030-35798-6},
  language = {en}
}

@incollection{huffman_integrated_2020,
  title = {Integrated {{Multi-satellite Retrievals}} for the {{Global Precipitation Measurement}} ({{GPM}}) {{Mission}} ({{IMERG}})},
  booktitle = {Satellite {{Precipitation Measurement}}},
  author = {Huffman, George J. and Bolvin, David T. and Braithwaite, Dan and Hsu, Kuo-Lin and Joyce, Robert J. and Kidd, Christopher and Nelkin, Eric J. and Sorooshian, Soroosh and Stocker, Erich F. and Tan, Jackson and Wolff, David B. and Xie, Pingping},
  year = 2020,
  volume = {67},
  pages = {343--353},
  address = {Cham},
  doi = {10.1007/978-3-030-24568-9_19},
  isbn = {978-3-030-24567-2 978-3-030-24568-9},
  language = {en}
}

@article{huffman_new_2023,
  title = {The {{New Version}} 3.2 {{Global Precipitation Climatology Project}} ({{GPCP}}) {{Monthly}} and {{Daily Precipitation Products}}},
  author = {Huffman, George J. and Adler, Robert F. and Behrangi, Ali and Bolvin, David T. and Nelkin, Eric J. and Gu, Guojun and Ehsani, Mohammad Reza},
  year = 2023,
  month = oct,
  journal = {Journal of Climate},
  volume = {36},
  number = {21},
  pages = {7635--7655},
  issn = {0894-8755, 1520-0442},
  doi = {10.1175/JCLI-D-23-0123.1},
  chapter = {Journal of Climate},
  language = {EN}
}

@article{jones_first_1999,
  title = {First- and {{Second-Order Conservative Remapping Schemes}} for {{Grids}} in {{Spherical Coordinates}}},
  author = {Jones, Philip W.},
  year = 1999,
  month = sep,
  journal = {Monthly Weather Review},
  volume = {127},
  number = {9},
  pages = {2204--2210},
  issn = {1520-0493, 0027-0644},
  doi = {10.1175/1520-0493(1999)127<2204:FASOCR>2.0.CO;2},
  chapter = {Monthly Weather Review},
  language = {EN}
}

@article{jong_increases_2023,
  title = {Increases in Extreme Precipitation over the {{Northeast United States}} Using High-Resolution Climate Model Simulations},
  author = {Jong, Bor-Ting and Delworth, Thomas L. and Cooke, William F. and Tseng, Kai-Chih and Murakami, Hiroyuki},
  year = 2023,
  month = mar,
  journal = {npj Climate and Atmospheric Science},
  volume = {6},
  number = {1},
  pages = {1--9},
  issn = {2397-3722},
  doi = {10.1038/s41612-023-00347-w},
  copyright = {2023 The Author(s)},
  language = {en}
}

@article{kooperman_robust_2016,
  title = {Robust Effects of Cloud Superparameterization on Simulated Daily Rainfall Intensity Statistics across Multiple Versions of the {{Community Earth System Model}}},
  author = {Kooperman, Gabriel J. and Pritchard, Michael S. and Burt, Melissa A. and Branson, Mark D. and Randall, David A.},
  year = 2016,
  journal = {Journal of Advances in Modeling Earth Systems},
  volume = {8},
  number = {1},
  pages = {140--165},
  issn = {1942-2466},
  doi = {10.1002/2015MS000574},
  copyright = {\copyright{} 2016. The Authors.},
  language = {en}
}

@article{kopparla_improved_2013,
  title = {Improved Simulation of Extreme Precipitation in a High-Resolution Atmosphere Model},
  author = {Kopparla, Pushkar and Fischer, Erich M. and Hannay, C{\'e}cile and Knutti, Reto},
  year = 2013,
  journal = {Geophysical Research Letters},
  volume = {40},
  number = {21},
  pages = {5803--5808},
  issn = {1944-8007},
  doi = {10.1002/2013GL057866},
  copyright = {\copyright 2013. American Geophysical Union. All Rights Reserved.},
  language = {en}
}

@misc{kyrouac_surface_2026,
  title = {Surface {{Meteorological Instrumentation}} ({{MET}}), 2004-11-01 to 2014-07-06, {{Tropical Western Pacific}} ({{TWP}}), {{Central Facility}}, {{Manus I}}., {{PNG}} ({{C1}})},
  author = {Kyrouac, Jenni and Shi, Yan and Tuftedal, Matt},
  year = 2026,
  month = feb,
  doi = {10.5439/1786358}
}

@article{lin_resolution_2023,
  title = {On the {{Resolution Sensitivity}} of {{Equatorial Precipitation}} in a {{GFDL Global Atmospheric Model}}},
  author = {Lin, Pu and Ming, Yi and Robinson, Thomas},
  year = 2023,
  journal = {Journal of Advances in Modeling Earth Systems},
  volume = {15},
  number = {10},
  pages = {e2022MS003300},
  issn = {1942-2466},
  doi = {10.1029/2022MS003300},
  copyright = {\copyright{} 2023 The Authors. This article has been contributed to by U.S. Government employees and their work is in the public domain in the USA.},
  language = {en}
}

@article{lovejoy_generalized_1985,
  title = {Generalized {{Scale Invariance}} in the {{Atmosphere}} and {{Fractal Models}} of {{Rain}}},
  author = {Lovejoy, S. and Schertzer, D.},
  year = 1985,
  journal = {Water Resources Research},
  volume = {21},
  number = {8},
  pages = {1233--1250},
  issn = {1944-7973},
  doi = {10.1029/WR021i008p01233},
  copyright = {Copyright 1985 by the American Geophysical Union.},
  language = {en}
}

@article{lovejoy_remarkable_2008,
  title = {The Remarkable Wide Range Spatial Scaling of {{TRMM}} Precipitation},
  author = {Lovejoy, S. and Schertzer, D. and Allaire, V. C.},
  year = 2008,
  month = oct,
  journal = {Atmospheric Research},
  volume = {90},
  number = {1},
  pages = {10--32},
  issn = {0169-8095},
  doi = {10.1016/j.atmosres.2008.02.016}
}

@article{lovejoy_review_2023,
  title = {Review Article: {{Scaling}}, Dynamical Regimes, and Stratification. {{How}} Long Does Weather Last? {{How}} Big Is a Cloud?},
  shorttitle = {Review Article},
  author = {Lovejoy, Shaun},
  year = 2023,
  month = aug,
  journal = {Nonlinear Processes in Geophysics},
  volume = {30},
  number = {3},
  pages = {311--374},
  issn = {1023-5809},
  doi = {10.5194/npg-30-311-2023},
  language = {English}
}

@article{mandelbrot_intermittent_1974,
  title = {Intermittent Turbulence in Self-Similar Cascades: Divergence of High Moments and Dimension of the Carrier},
  shorttitle = {Intermittent Turbulence in Self-Similar Cascades},
  author = {Mandelbrot, Benoit B.},
  year = 1974,
  month = jan,
  journal = {Journal of Fluid Mechanics},
  volume = {62},
  number = {2},
  pages = {331--358},
  issn = {1469-7645, 0022-1120},
  doi = {10.1017/S0022112074000711},
  language = {en}
}

@article{marani_nonpowerlawscale_2005,
  title = {Non-Power-Law-Scale Properties of Rainfall in Space and Time},
  author = {Marani, Marco},
  year = 2005,
  journal = {Water Resources Research},
  volume = {41},
  number = {8},
  issn = {1944-7973},
  doi = {10.1029/2004WR003822},
  language = {en}
}

@article{martinez-villalobos_why_2019,
  title = {Why {{Do Precipitation Intensities Tend}} to {{Follow Gamma Distributions}}?},
  author = {{Martinez-Villalobos}, Cristian and Neelin, J. David},
  year = 2019,
  month = nov,
  journal = {Journal of the Atmospheric Sciences},
  volume = {76},
  number = {11},
  pages = {3611--3631},
  issn = {0022-4928, 1520-0469},
  doi = {10.1175/JAS-D-18-0343.1},
  chapter = {Journal of the Atmospheric Sciences},
  language = {EN}
}

@article{masunaga_interproduct_2019,
  title = {Inter-Product Biases in Global Precipitation Extremes},
  author = {Masunaga, Hirohiko and Schr{\"o}der, Marc and Furuzawa, Fumie A and Kummerow, Christian and Rustemeier, Elke and Schneider, Udo},
  year = 2019,
  month = dec,
  journal = {Environmental Research Letters},
  volume = {14},
  number = {12},
  pages = {125016},
  issn = {1748-9326},
  doi = {10.1088/1748-9326/ab5da9},
  language = {en}
}

@article{mitchell_overview_1976,
  title = {An Overview of Climatic Variability and Its Causal Mechanisms},
  author = {Mitchell, J. Murray},
  year = 1976,
  month = dec,
  journal = {Quaternary Research},
  volume = {6},
  number = {4},
  pages = {481--493},
  issn = {0033-5894},
  doi = {10.1016/0033-5894(76)90021-1}
}

@misc{niedorf_girafe_2024,
  title = {{{GIRAFE}} v1: {{CM SAF Global Interpolated RAinFall Estimation}} Version 1},
  shorttitle = {{{GIRAFE}} V1},
  author = {Niedorf, Anja and Finkensieper, Stephan and Konrad, Hannes and Roca, R{\'e}my and Marc, Schr{\"o}der and Cloch{\'e}, Sophie and Panegrossi, Giulia and San{\`o}, Paolo and Kidd, Christopher and Juc{\'a} Oliveira, R{\^o}mulo Augusto and Fennig, Karsten and Sikorski, Thomas and {Penning de Vries}, Marloes and Radovan, Ana and Dietzsch, Felix and Pondrom, Marc and Selbach, Nathalie and Hollmann, Rainer},
  year = 2024,
  month = apr,
  pages = {2.9 GiB},
  doi = {10.5676/EUM_SAF_CM/GIRAFE/V001}
}

@article{nikumbh_does_2024,
  title = {Does {{Increasing Horizontal Resolution Improve}} the {{Simulation}} of {{Intense Tropical Rainfall}} in {{GFDL}}'s {{AM4 Model}}?},
  author = {Nikumbh, Akshaya C. and Lin, Pu and Paynter, David and Ming, Yi},
  year = 2024,
  journal = {Geophysical Research Letters},
  volume = {51},
  number = {12},
  pages = {e2023GL106708},
  issn = {1944-8007},
  doi = {10.1029/2023GL106708},
  copyright = {\copyright{} 2024. The Author(s).},
  language = {en}
}

@article{nishant_impact_2022,
  title = {Impact of {{Higher Spatial Resolution}} on {{Precipitation Properties Over Australia}}},
  author = {Nishant, Nidhi and Sherwood, Steven and Prasad, Abhnil and Ji, Fei and Singh, Ajit},
  year = 2022,
  journal = {Geophysical Research Letters},
  volume = {49},
  number = {23},
  pages = {e2022GL100717},
  issn = {1944-8007},
  doi = {10.1029/2022GL100717},
  copyright = {\copyright{} 2022. The Authors.},
  language = {en}
}

@article{nystuen_relative_1999,
  title = {Relative {{Performance}} of {{Automatic Rain Gauges}} under {{Different Rainfall Conditions}}},
  author = {Nystuen, Jeffrey A.},
  year = 1999,
  month = aug,
  journal = {Journal of Atmospheric and Oceanic Technology},
  volume = {16},
  number = {8},
  pages = {1025--1043},
  issn = {0739-0572, 1520-0426},
  doi = {10.1175/1520-0426(1999)016<1025:RPOARG>2.0.CO;2},
  chapter = {Journal of Atmospheric and Oceanic Technology},
  language = {EN}
}

@article{olsson_effect_1999,
  title = {Effect of Spatial Averaging on Temporal Statistical and Scaling Properties of Rainfall},
  author = {Olsson, Jonas and Singh, Vijay P. and Jinno, Kenji},
  year = 1999,
  journal = {Journal of Geophysical Research: Atmospheres},
  volume = {104},
  number = {D16},
  pages = {19117--19126},
  issn = {2156-2202},
  doi = {10.1029/1999JD900271},
  copyright = {Copyright 1999 by the American Geophysical Union.},
  language = {en}
}

@article{over_spacetime_1996,
  title = {A Space-Time Theory of Mesoscale Rainfall Using Random Cascades},
  author = {Over, Thomas M. and Gupta, Vijay K.},
  year = 1996,
  journal = {Journal of Geophysical Research: Atmospheres},
  volume = {101},
  number = {D21},
  pages = {26319--26331},
  issn = {2156-2202},
  doi = {10.1029/96JD02033},
  language = {en}
}

@article{over_statistical_1994,
  title = {Statistical {{Analysis}} of {{Mesoscale Rainfall}}: {{Dependence}} of a {{Random Cascade Generator}} on {{Large-Scale Forcing}}},
  shorttitle = {Statistical {{Analysis}} of {{Mesoscale Rainfall}}},
  author = {Over, Thomas M. and Gupta, Vijay K.},
  year = 1994,
  month = dec,
  journal = {Journal of Applied Meteorology and Climatology},
  volume = {33},
  number = {12},
  pages = {1526--1542},
  issn = {1520-0450, 0894-8763},
  doi = {10.1175/1520-0450(1994)033<1526:SAOMRD>2.0.CO;2},
  chapter = {Journal of Applied Meteorology and Climatology},
  language = {EN}
}

@article{pendergrass_climatological_2017,
  title = {Climatological {{Characteristics}} of {{Typical Daily Precipitation}}},
  author = {Pendergrass, Angeline G. and Deser, Clara},
  year = 2017,
  month = aug,
  journal = {Journal of Climate},
  volume = {30},
  number = {15},
  pages = {5985--6003},
  issn = {0894-8755, 1520-0442},
  doi = {10.1175/JCLI-D-16-0684.1},
  chapter = {Journal of Climate},
  language = {EN}
}

@article{pendergrass_two_2014,
  title = {Two {{Modes}} of {{Change}} of the {{Distribution}} of {{Rain}}},
  author = {Pendergrass, Angeline G. and Hartmann, Dennis L.},
  year = 2014,
  month = nov,
  journal = {Journal of Climate},
  volume = {27},
  number = {22},
  pages = {8357--8371},
  issn = {0894-8755, 1520-0442},
  doi = {10.1175/JCLI-D-14-00182.1},
  chapter = {Journal of Climate},
  language = {EN}
}

@article{rajulapati_perils_2021,
  title = {The {{Perils}} of {{Regridding}}: {{Examples Using}} a {{Global Precipitation Dataset}}},
  shorttitle = {The {{Perils}} of {{Regridding}}},
  author = {Rajulapati, Chandra Rupa and Papalexiou, Simon Michael and Clark, Martyn P. and Pomeroy, John W.},
  year = 2021,
  month = nov,
  journal = {Journal of Applied Meteorology and Climatology},
  volume = {60},
  number = {11},
  pages = {1561--1573},
  issn = {1558-8424, 1558-8432},
  doi = {10.1175/JAMC-D-20-0259.1},
  chapter = {Journal of Applied Meteorology and Climatology},
  language = {EN}
}

@article{roca_frogs_2019,
  title = {{{FROGS}}: A Daily 1{$^\circ$} \texttimes{} 1{$^\circ$} Gridded Precipitation Database of Rain Gauge, Satellite and Reanalysis Products},
  shorttitle = {{{FROGS}}},
  author = {Roca, R{\'e}my and Alexander, Lisa V. and Potter, Gerald and Bador, Margot and Juc{\'a}, R{\^o}mulo and Contractor, Steefan and Bosilovich, Michael G. and Cloch{\'e}, Sophie},
  year = 2019,
  month = jul,
  journal = {Earth System Science Data},
  volume = {11},
  number = {3},
  pages = {1017--1035},
  issn = {1866-3508},
  doi = {10.5194/essd-11-1017-2019},
  language = {English}
}

@article{schertzer_generalised_1985,
  title = {Generalised Scale Invariance in Turbulent Phenomena},
  author = {Schertzer, D. and Lovejoy, S.},
  year = 1985,
  journal = {PhysicoChemical Hydrodynamics},
  volume = {6},
  number = {5-6},
  pages = {623--635},
  issn = {0191-9059},
  language = {en}
}

@article{schumacher_stratiform_2003,
  title = {Stratiform {{Rain}} in the {{Tropics}} as {{Seen}} by the {{TRMM Precipitation Radar}}},
  author = {Schumacher, Courtney and Houze, Robert A.},
  year = 2003,
  month = jun,
  journal = {Journal of Climate},
  volume = {16},
  number = {11},
  pages = {1739--1756},
  issn = {0894-8755, 1520-0442},
  doi = {10.1175/1520-0442(2003)016<1739:SRITTA>2.0.CO;2},
  chapter = {Journal of Climate},
  language = {EN}
}

@article{slingo_ambitious_2022,
  title = {Ambitious Partnership Needed for Reliable Climate Prediction},
  author = {Slingo, Julia and Bates, Paul and Bauer, Peter and Belcher, Stephen and Palmer, Tim and Stephens, Graeme and Stevens, Bjorn and Stocker, Thomas and Teutsch, Georg},
  year = 2022,
  month = jun,
  journal = {Nature Climate Change},
  volume = {12},
  number = {6},
  pages = {499--503},
  issn = {1758-6798},
  doi = {10.1038/s41558-022-01384-8},
  copyright = {2022 Springer Nature Limited},
  language = {en}
}

@article{tan_performance_2017,
  title = {Performance of {{IMERG}} as a {{Function}} of {{Spatiotemporal Scale}}},
  author = {Tan, Jackson and Petersen, Walter A. and Kirstetter, Pierre-Emmanuel and Tian, Yudong},
  year = 2017,
  month = feb,
  journal = {Journal of Hydrometeorology},
  volume = {18},
  number = {2},
  pages = {307--319},
  issn = {1525-7541, 1525-755X},
  doi = {10.1175/JHM-D-16-0174.1},
  chapter = {Journal of Hydrometeorology},
  language = {EN}
}

@article{trenberth_intermittency_2017,
  title = {Intermittency in {{Precipitation}}: {{Duration}}, {{Frequency}}, {{Intensity}}, and {{Amounts Using Hourly Data}}},
  shorttitle = {Intermittency in {{Precipitation}}},
  author = {Trenberth, Kevin E. and Zhang, Yongxin and Gehne, Maria},
  year = 2017,
  month = may,
  journal = {Journal of Hydrometeorology},
  volume = {18},
  number = {5},
  pages = {1393--1412},
  issn = {1525-7541, 1525-755X},
  doi = {10.1175/JHM-D-16-0263.1},
  chapter = {Journal of Hydrometeorology},
  language = {EN}
}

@article{trenberth_nearglobal_2018,
  title = {Near-{{Global Covariability}} of {{Hourly Precipitation}} in {{Space}} and {{Time}}},
  author = {Trenberth, Kevin E. and Zhang, Yongxin},
  year = 2018,
  month = apr,
  journal = {Journal of Hydrometeorology},
  volume = {19},
  number = {4},
  pages = {695--713},
  issn = {1525-7541, 1525-755X},
  doi = {10.1175/JHM-D-17-0238.1},
  chapter = {Journal of Hydrometeorology},
  language = {EN}
}

@article{veneziano_imperfect_2006,
  title = {Imperfect Scaling of Time and Space--Time Rainfall},
  author = {Veneziano, Daniele and Furcolo, Pierluigi and Iacobellis, Vito},
  year = 2006,
  month = may,
  journal = {Journal of Hydrology},
  series = {Hydrofractals '03},
  volume = {322},
  number = {1},
  pages = {105--119},
  issn = {0022-1694},
  doi = {10.1016/j.jhydrol.2005.02.044}
}

@article{venugopal_evidence_1999,
  title = {Evidence of Dynamic Scaling in Space-Time Rainfall},
  author = {Venugopal, V. and {Foufoula-Georgiou}, Efi and Sapozhnikov, Victor},
  year = 1999,
  journal = {Journal of Geophysical Research: Atmospheres},
  volume = {104},
  number = {D24},
  pages = {31599--31610},
  issn = {2156-2202},
  doi = {10.1029/1999JD900437},
  copyright = {Copyright 1999 by the American Geophysical Union.},
  language = {en}
}

@article{wood_distribution_2011,
  title = {The {{Distribution}} of {{Cloud Horizontal Sizes}}},
  author = {Wood, Robert and Field, Paul R.},
  year = 2011,
  month = sep,
  journal = {Journal of Climate},
  volume = {24},
  number = {18},
  pages = {4800--4816},
  issn = {0894-8755, 1520-0442},
  doi = {10.1175/2011JCLI4056.1},
  chapter = {Journal of Climate},
  language = {EN}
}

@article{xie_reprocessed_2017,
  title = {Reprocessed, {{Bias-Corrected CMORPH Global High-Resolution Precipitation Estimates}} from 1998},
  author = {Xie, Pingping and Joyce, Robert and Wu, Shaorong and Yoo, Soo-Hyun and Yarosh, Yelena and Sun, Fengying and Lin, Roger},
  year = 2017,
  month = jun,
  journal = {Journal of Hydrometeorology},
  volume = {18},
  number = {6},
  pages = {1617--1641},
  issn = {1525-7541, 1525-755X},
  doi = {10.1175/JHM-D-16-0168.1},
  chapter = {Journal of Hydrometeorology},
  language = {EN}
}

@article{xu_properties_2012,
  title = {Properties of Deep Convection in Tropical Continental, Monsoon, and Oceanic Rainfall Regimes},
  author = {Xu, Weixin and Zipser, Edward J.},
  year = 2012,
  journal = {Geophysical Research Letters},
  volume = {39},
  number = {7},
  issn = {1944-8007},
  doi = {10.1029/2012GL051242},
  copyright = {Copyright 2012 by the American Geophysical Union},
  language = {en}
}

@misc{zhu_weighing_2026,
  title = {Weighing {{Bucket Precipitation Gauge}} ({{WBPLUVIO2}}), 2016-11-22 to 2024-03-31, {{Southern Great Plains}} ({{SGP}}), {{Central Facility}}, {{Lamont}}, {{OK}} ({{C1}})},
  author = {Zhu, Zeen and Wang, Die and Jane, Mary and Cromwell, Erol and Sturm, Matthew and Irving, Kenneth and Delamere, Jennifer},
  year = 2026,
  month = feb,
  doi = {10.5439/1338194}
}

\end{document}


\maketitle

%
%
%
%
%
%
\section*{Supplementary Materials}
\begin{figure*}[b]
\noindent\includegraphics[width=5.5in,angle=0]{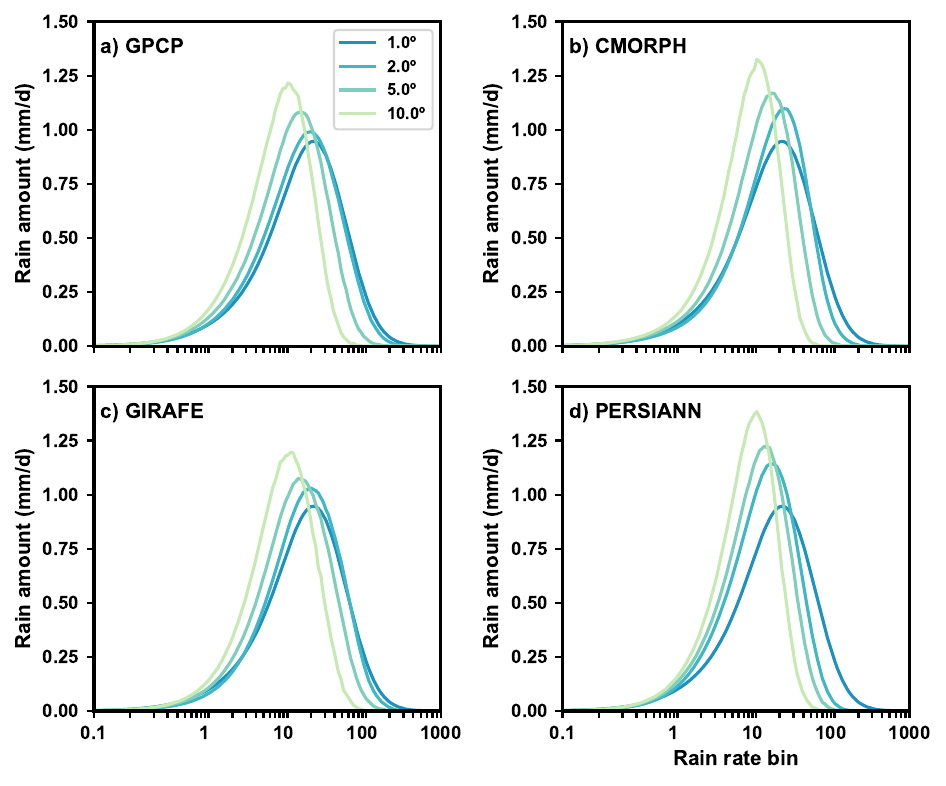}\\
\caption{Precipitation amount distributions at different spatial resolutions for a) GPCP, b) CMORPH, c) GIRAFE and d)PERSIANN. Each distribution was created from the native (1º) and regridded (2º,5º and 10º) spatial resolutions from each dataset. The different colors correspond to different spatial resolutions.}\label{figS2}
\end{figure*}
\begin{figure*}
\noindent\includegraphics[width=5.5in,angle=0]{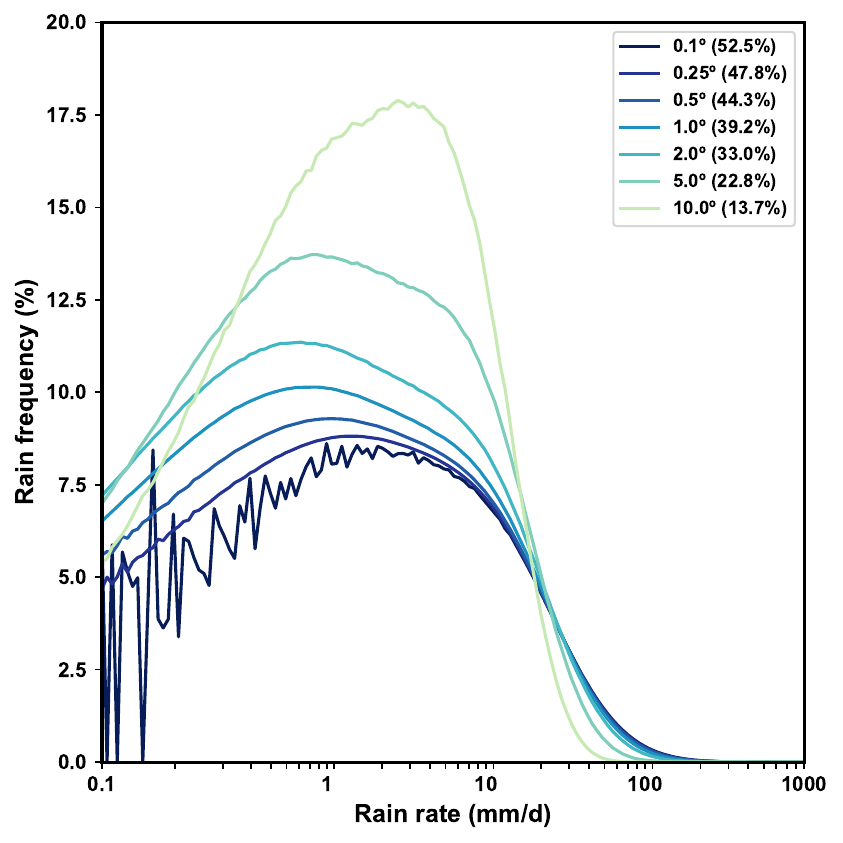}\\
\caption{Precipitation frequency distributions at different spatial resolutions. Each distribution was created from the native (0.1º) and regridded (0.25º,0.5º,1º,2º,5º and 10º) spatial resolutions from IMERG. The different colors correspond to different spatial resolutions. The dry day frequency associated with each spatial resolution is provided in the legend.}\label{figS1}
\end{figure*}